\newcommand*{\ATLASLATEXPATH}{}
\documentclass[cernpreprint,UKenglish,txfonts,texlive=2011]{\ATLASLATEXPATH atlasdoc}
\pdfoutput=1

\usepackage[biblatex=false]{\ATLASLATEXPATH atlaspackage}

\usepackage{\ATLASLATEXPATH atlascontribute}

\usepackage{\ATLASLATEXPATH atlasphysics}

\usepackage{rotating}

\usepackage{HeavyIonPhotonPaper2015-defs}

\hypersetup{pdftitle={ATLAS draft},pdfauthor={The ATLAS Collaboration}}
\AtlasJournal{Phys.\ Rev.\ C}
\PreprintIdNumber{CERN-PH-EP-2015-142}
\AtlasDate{\today}
\AtlasTitle{Centrality, rapidity and transverse momentum dependence of isolated prompt photon production in lead-lead collisions at \energy\ measured with the ATLAS detector}
\author{The ATLAS Collaboration}

\AtlasAbstract{
  Prompt photon production in \mbox{$\sqrt{s_{NN}}=2.76$ TeV}
  Pb+Pb collisions has been measured by the ATLAS experiment at
  the Large Hadron Collider using data collected in 2011 with an
  integrated luminosity of 0.14 $\mathrm{nb}^{-1}$.  Inclusive photon
  yields, scaled by the mean nuclear thickness function, are presented
  as a function of collision centrality and transverse momentum in two
  pseudorapidity intervals, $|\eta|<1.37$ and $1.52 \leq |\eta|<2.37$.
  The scaled yields in the two pseudorapidity intervals, as well as
  the ratios of the forward yields to those at midrapidity, are
  compared to the expectations from next-to-leading order perturbative
  QCD calculations.  The measured cross sections
  agree well with the predictions for proton-proton collisions within
  statistical and systematic uncertainties.  Both the yields and
  ratios are also compared to two other pQCD calculations, one which
  uses the isospin content appropriate to colliding lead nuclei, and
  another which includes nuclear modifications to the
  nucleon parton distribution functions.  }

\begin{document}

\maketitle

\tableofcontents

\section{Introduction}
\label{sec:intro}

Prompt photons
are an important probe for the study of the hot, dense matter formed in the high-energy collision of heavy ions.
Being colorless, they are transparent to the subsequent evolution of the matter and probe the very initial stages
of the collision.
Their production rates are therefore expected to be directly sensitive to the overall thickness of the colliding nuclear matter.
The rates are also expected to be sensitive to modifications of the partonic structure of nucleons bound in a nucleus,
which are implemented as nuclear modifications~\cite{Eskola:2009uj,Hirai:2007sx,deFlorian:2003qf}
to the parton distribution functions (PDF) measured in deep-inelastic lepton-proton
and proton-proton ($\pp$) scattering experiments.
These effects include nuclear shadowing (the depletion of the parton
densities at low Bjorken $x$), anti-shadowing (an enhancement at
moderate $x$), and the EMC effect~\cite{Aubert:1983xm}.  Photon rates
are also sensitive to final-state interactions in the hot and dense
medium, via the conversion of high energy quarks and gluons into
photons through rescattering.
This is predicted to lead to an increased photon production
rate relative to standard
expectations~\cite{Fries:2002kt,Turbide:2005fk}.

Prompt photons
have two primary sources.  The first is direct emission, which
proceeds at leading order via quark-gluon Compton scattering $qg
\rightarrow q\gamma$ or quark-antiquark annihilation $\overline{q}q
\rightarrow g\gamma$.
The second is the fragmentation contribution from the production of hard
photons during parton fragmentation.
At leading order in perturbative quantum chromodynamics (pQCD) calculations,
there is a meaningful distinction between the direct emission
and fragmentation, but at higher orders
the two cannot be unambiguously separated.
In order to suppress the large
background of nonprompt photons originating from the
decays of neutral mesons in jets,
as well as fragmentation photons,
an isolation criterion is applied, both in measurements and calculations,
to the transverse energy
contained within a cone of well-defined size
around the photon direction~\cite{Catani:2002ny}.
The isolation transverse energy requirement can be applied as a fraction of the photon
transverse energy, or as a constant transverse energy threshold.
In either case, these requirements can be applied consistently to pQCD calculations so that
prompt photon rates can be calculated reliably, as the isolation criterion
naturally cuts off the collinear divergence of the fragmentation
contribution~\cite{Catani:2002ny}.

Prompt photon rates have been measured extensively in both fixed target and collider experiments.
Fixed target experiments include WA70~\cite{WA70}, UA6~\cite{UA6} and E706~\cite{Apanasevich:2004dr}, and cover the range $\sqrt{s}= 23 \ndash 38.8$~\GeV.
In collider experiments, measurements were performed for
proton-proton collisions at the CERN Intersecting Storage Rings ($\pp$, \mbox{$\sqrt{s}=24 \ndash 62.4$~\GeV} )~\cite{Anassontzis:1982gm,Angelis:1989zv},
and BNL Relativistic Heavy Ion Collider ($\pp$ at $\sqrt{s}=$200~\GeV )~\cite{Adare:2012yt,Abelev:2009hx},
and for proton-antiproton collisions
at the CERN Super Proton Synchrotron ($\pbarp$, \mbox{$\sqrt{s}=546 \ndash 630$}~\GeV)~\cite{Albajar:1988im,Alitti:1992hn}
and at the Fermilab Tevatron ($\pbarp$, \mbox{$\sqrt{s}=0.63 \ndash 1.96$}~\TeV)~\cite{Abbott:1999kd,Acosta:2002ya,Abazov:2005wc,Aaltonen:2009ty}.
At the CERN Large Hadron Collider (LHC), \atlas~\cite{Aad:2010sp,Aad:2011tw,Aad:2013zba} and CMS~\cite{Khachatryan:2010fm,Chatrchyan:2011ue}
have measured isolated prompt photons in $\pp$ collisions at \mbox{$\sqrt{s}=7$}~\TeV.
In most cases, good agreement has been found with pQCD predictions
at next-to-leading order (NLO),
which are typically calculated using the \jetphox\ package~\cite{Catani:2002ny,Aurenche:2006vj}.
In lower-energy heavy-ion collisions,
the WA98 experiment
observed direct photons in lead-lead ($\PbPb$) collisions at
\mbox{$\sqrt{s_{\mathrm{NN}}} = 17.3$}~\GeV\ ~\cite{Aggarwal:2000th}, and
the PHENIX experiment performed  measurements of direct photon rates in gold-gold collisions at
$\sqrt{s_{\mathrm{NN}}} = 200$~\GeV\ \cite{Adler:2005ig,Afanasiev:2012dg}.

A variable often used to characterize the modification of
rates of hard processes
in a nuclear environment is the nuclear modification factor
\begin{equation}
R_{\mathrm{AA}} = \frac{(1/N_\mathrm{evt}) dN_\mathrm{X}/d\pT }{\meanTAA d\sigma^{pp}_\mathrm{X}/d\pT },
\end{equation}
where $dN_X/d\pT$ is the yield of objects X produced in a $\pT$ interval,
$N_{\mathrm{evt}}$ is the number of sampled minimum-bias events,
$\TAA$ is the mean nuclear thickness function
(defined as the mean number of binary collisions divided by the total inelastic
nucleon-nucleon ($NN$) cross section) and $d\sigma^{pp}_\mathrm{X} / d\pT$ is the cross section of process $X$ in $\pp$ collisions for the same $\pT$ interval.
With this formula,
one can make straightforward comparisons of yields in heavy-ion collisions,
normalized by the flux of initial-state partons,
to those measured in \pp\ data, or calculated in pQCD.
CMS performed the first measurement of isolated prompt photon rates in
both $\PbPb$ and $\pp$ collisions at $\sqrt{s}=$2.76~\TeV\ up to a photon transverse energy
\mbox{$\ET=80$}~\GeV\ within \mbox{$|\eta|<1.44$}~\cite{Chatrchyan:2012vq}.
This measurement observed prompt, isolated photon rates consistent with $R_{\mathrm{AA}}=1$
for all collision impact parameters and $\ET$ ranges considered, and good agreement of the
data with \jetphox\ calculations.

This \ptype\ presents isolated prompt photon yields, scaled by the mean
nuclear thickness to derive effective cross sections,
measured in $\PbPb$ collisions  with the
\atlas\ detector, making use of its large-acceptance, longitudinally segmented
calorimeter system.
The effect of the underlying event (UE) on the photon energy and shower shape
is corrected on an event-by-event basis.
Photon yields are
measured over two ranges in the pseudorapidity of the photon,
$|\eta|<1.37$ (central) and $1.52 \leq |\eta| < 2.37$ (forward),
and for photon transverse momenta in the interval $22 \leq \pT < 280$~$\GeV$.
Comparisons of the yields
with NLO pQCD calculations
are also presented from \jetphox\ 1.3~\cite{Aurenche:2006vj},
in three configurations:
\pp\ collisions, \PbPb\ collisions (i.e. with the correct total isospin),
and \PbPb\ after incorporating the EPS09
nuclear modification factors to the nucleon PDFs~\cite{Eskola:2009uj},
derived from experimental data of lepton and proton scattering on nuclei.
The ratios of the yields in the forward $\eta$ region to those in the
central $\eta$ region ($\RFC$) are also presented.

\section{ATLAS detector}
\label{sec:detector}

The \atlas\ detector
comprises three major subsystems: the inner detector, the calorimeter system,
and the muon spectrometer.
It is described in detail in Ref.~\cite{Aad:2008zzm}.

The inner detector is comprised of
the pixel detector, the semiconductor tracker (SCT)
and the transition radiation tracker (TRT), which cover
the full azimuthal range and pseudorapidities\footnote{
\atlas\ uses a
right-handed coordinate system with its origin at the nominal
interaction point (IP) in the center of the detector and the
$z$-axis along the beam pipe. The $x$-axis points from the IP to the
center of the LHC ring, and the $y$ axis points upward. Cylindrical
coordinates $(r,\phi)$ are used in the transverse plane, $\phi$
being the azimuthal angle around the beam pipe. The pseudorapidity
is defined in terms of the polar angle $\theta$ as
$\eta=-\ln\tan(\theta/2)$.
}
$|\eta|<2.5$, except for the TRT, which covers $|\eta|<2$.
The muon spectrometer measures muons over $|\eta|<2.7$ with a combination of
monitored drift tubes and cathode strip chambers.

The \atlas\ calorimeter is the primary subsystem used for the measurement presented here.
It is a large-acceptance, longitudinally segmented sampling calorimeter covering $|\eta|<4.9$ with electromagnetic (EM) and hadronic sections.
The EM section is a lead/liquid-argon
sampling calorimeter with an accordion-shaped geometry.
It is divided into a barrel region, covering $|\eta|<1.475$, and two endcap regions, covering
$1.375<|\eta|<3.2$.
The EM calorimeter has three primary sections, longitudinal in shower depth,
called ``layers,'' to fully contain photon showers in the range of interest for this
analysis.  The first sampling layer is
3 to 5 radiation lengths deep and is segmented into fine strips of size $\Delta \eta=0.003-0.006$ (depending on
$\eta$), which allows the discrimination of photons from the two-photon decays of $\pi^0$ and $\eta$ mesons.
The second layer is 17 radiation lengths thick, sampling most
of an electromagnetic shower,
and has cells of size $ \dedp = 0.025 \times 0.025$.
The third layer has a material depth ranging from 4 to 15 radiation lengths and is used to correct for the leakage beyond the first two layers for
high-energy electromagnetic showers.
The total material in front of the electromagnetic
calorimeter ranges from 2.5 to 6 radiation lengths depending on pseudorapidity,
except in the transition region between the barrel and endcap regions ($1.37\leq |\eta| < 1.52$),
in which the material is up to 11.5 radiation lengths (for which reason this transition
region is excluded from this analysis).
In front of the strip layer, a presampler is used to correct for energy loss in front of the calorimeter within
the region $|\eta|<1.8$.
In test beam environments and in typical \pp\ collisions,
the photon energy resolution
is found to have a sampling term of 10--17\% $ / \sqrt{E[\mathrm{GeV}]} $.
Above 200~$\GeV$ the global constant term in the photon energy resolution,
estimated to be $1.2\%\pm 0.6\% $ ($1.8\% \pm 0.6\%$) in the barrel (endcap) region
for \pp\ data at $\sqrt{s}=7$~\TeV, starts to dominate~\cite{ATLAS-CONF-2010-005}.
The hadronic calorimeter section is located
outside the electromagnetic calorimeter.  Within $|\eta|<1.7$, it is a
sampling calorimeter of steel and scintillator tiles, with a depth of
7.4 hadronic interaction lengths.

The \atlas\ zero-degree calorimeters (ZDCs) are used for minimum-bias event triggering.
They detect forward-going neutral particles with $|\eta|>8.3$.
The minimum-bias trigger scintillators (MBTS) detect charged particles in the interval $2.1<|\eta|<3.9$
using two sets of 16 counters positioned at $z = \pm 3.6\,$m.  They are used for event selection.
The forward calorimeter (FCal) is
used to determine the ``centrality'' of the collision,
which can be related to geometric parameters such as the number of participating nucleons
or the number of binary collisions~\cite{Miller:2007ri}.
The FCal has three layers in the longitudinal direction, one electromagnetic
and two hadronic, covering $3.1<|\eta|<4.9$.
The FCal electromagnetic and hadronic modules are composed of
copper and tungsten absorbers, respectively, with liquid
argon as the active medium, which together provide ten interaction
lengths of material.

The sample of events used in this analysis was collected
using the first-level calorimeter trigger~\cite{Achenbach:2008zzb}.
This is a hardware trigger that sums the electromagnetic energy in
towers of size $\Delta \eta \times \Delta \phi = 0.1 \times 0.1$.
A sliding window of size $0.2 \times 0.2$ was used to find electromagnetic clusters by
searching for local energy
maxima and keeping only those clusters with energy in two
adjacent towers (i.e. regions with a size of either $0.2\times 0.1$ or $0.1 \times 0.2$) exceeding
a threshold.
The trigger used for the present measurement had a threshold of
16~\GeV\ transverse energy.

\section{Collision data selection}
\label{sect:collision}
The data sample analyzed in this \ptype\ corresponds to an integrated luminosity of
$\Lint = 0.14$ $\mathrm{nb}^{-1}$ \PbPb\ collisions at \energy\
collected during the 2011 LHC heavy-ion run.
After the trigger requirement,
events must satisfy a set of selection criteria.
To suppress backgrounds,
the relative time measured between the two MBTS counters is
required to be less than 5~ns,
and a primary vertex is required to be reconstructed in the inner detector.
Minimum-bias events were triggered in the same data samples
based on either a coincidence in the two ZDCs associated with a track in the inner detector, or
a total of at least 50~\GeV\ transverse energy
deposited in the full calorimeter system.
These events were also required to pass the same MBTS and vertex selections as the photon-triggered events.
To be consistent with the minimum-bias trigger selections,
a ZDC coincidence is also required for photon-triggered events with low FCal $\sumet$.

The centrality of each heavy-ion collision is determined using the total
transverse energy measured in the forward calorimeter ($3.2<|\eta|<4.9$),
at the electromagnetic scale, FCal $\sumet$.
The trigger and event selection were studied in detail in the 2010 $\PbPb$ data
sample~\cite{ATLAS:2011ah} and
$98\pm2\%$ of the total inelastic cross section was accepted.
The higher luminosity of the 2011 heavy-ion run necessitated a more
sophisticated trigger strategy, including more restrictive
triggers in the most peripheral events.
However, it was found that the FCal $\sumet$ distributions in 2011 data
match those measured in 2010 to a high degree of precision.
For this analysis, the FCal $\sumet$ distribution
was divided into four centrality intervals, covering the 0--10\%, 10--20\%, 20--40\% and 40--80\% most central events.
With this convention, the 0--10\% interval contains the events with
the largest forward transverse energy production, and the 40--80\%
interval the smallest.
The total number of minimum-bias events corresponding to the 0--80\% centrality interval
is $N_{\mathrm{evt}} = 7.93 \times 10^8$.

Quantities which describe the average geometric configuration of the
colliding nuclei
are calculated as described in Ref.~\cite{ATLAS:2011ag} using a Glauber Monte-Carlo calculation to describe the measured
minimum-bias FCal distribution.
Table~\ref{table:centrality} summarizes all of the centrality-related information used
in this analysis.
For each centrality interval,
the table specifies
the mean number of nucleons that interact at least once $\meanNpart$,
the mean number of binary collisions $\meanNcoll$,
and the mean value of the nuclear thickness function $\meanTAA$,
with their respective fractional uncertainties.
The uncertainty on the mean nuclear thickness function $\meanTAA=\meanNcoll/\sigmaNN$ is smaller than the corresponding
uncertainty on $\meanNcoll$, since the uncertainty on $\sigmaNN$ largely cancels in the ratio.
All of the uncertainties account for variations in the Glauber model parameters consistent with
the uncertainties about the nuclear wave function, as well as the uncertainty in the estimation of the measured
fraction of the total inelastic cross section.

Since the distribution of FCal $\sumet$ is different in events with high-$\pT$ photons
compared to minimum-bias events,
a weighting factor is applied to each simulated event
to make the simulated distributions
agree with the measured distributions.

\begin{table*}[t]
\begin{center}
\caption
{ Centrality bins used in this analysis, tabulating the percentage
  range, the average number of participants ($\langle \Npart\rangle$)
  and binary collisions ($\langle \Ncoll\rangle$), the mean nuclear
  thickness ($\meanTAA$) and the relative systematic uncertainty on
  these quantities.
\label{table:centrality}
}
\begin{tabular}{|c||c|c|c|c|c|c|}
\hline
Interval &  $\langle \Npart \rangle$ & $\frac{\delta \langle \Npart \rangle}{ \langle \Npart \rangle} $ & $\langle \Ncoll \rangle$ & $\frac{ \delta \langle \Ncoll \rangle}{ \langle \Ncoll \rangle}$ & $\langle \TAA \rangle$ [mb$^{-1}$]& $\frac{\delta \langle \TAA \rangle} { \langle \TAA \rangle}$  \\
\hline
\hline
0--10\% &  356.2 & 0.7\% & 1500.6 & 7.6\% & 23.4 & 1.6\% \\
10--20\% & 260.7 & 1.4\% & 923.3 & 7.3\% & 14.4 & 2.1\%\\
20--40\% & 157.8 & 2.4\% & 440.6 & 7.2\% & 6.9 & 3.5\%\\
40--80\% & 45.9 & 5.9\% & 77.8 & 9.1\% & 1.2 & 8.1\%\\
\hline
\end{tabular}
\end{center}
\end{table*}

\section{Simulated data samples}
\label{section:sim}
For the extraction of photon reconstruction and identification efficiencies,
the photon energy scale, and expected properties of
the isolation transverse energy distributions,
samples of events containing prompt photons were produced using
\PYTHIA\ 6.423~\cite{Sjostrand:2006za}
for $\pp$ collisions at $\sqrt{s}=2.76$~\TeV\ using the ATLAS AUET2B
set of tuned parameters
~\cite{ATL-PHYS-PUB-2011-009}.
Direct photons were simulated in
photon-jet events
divided into four sub-samples based on requiring a minimum $\pT$ for
the primary photon: $\pT> 17$~\GeV, $\pT> 35$~\GeV, $\pT> 70$~\GeV, and $\pT> 140$~\GeV.
The contribution of fragmentation photons was modeled using a
set of
simulated inclusive jet $\pp$ events, also using the same \PYTHIA\ 6 tune.
Each of these is required to have
a hard photon produced in the fragmentation of
jets produced with the \PYTHIA\ 6 hardness
scale, which controls the typical $\pT$ of the produced jets, ranging
from 17 to 560~\GeV.
Similar samples were also prepared using the \SHERPA\ generator~\cite{Gleisberg:2008ta}
using the CT10~\cite{Lai:2010vv} parton distribution functions, which include
both direct and fragmentation photon contributions.
These were used to check on the
generator dependence of the photon efficiency.
A large sample of \PYTHIA\ 6 inclusive-jet events, without the hard photon requirement, were utilized
to study the properties of background candidates.
For all generated samples, each event was fully simulated using
GEANT4~\cite{Agostinelli:2002hh,Aad:2010ah}.

Each simulated event is overlaid upon a real minimum-bias event from
experimental data, with the simulated event vertex placed
at the position of the measured vertex position.
By using minimum-bias data as the underlying-event model,
almost all features of the underlying event
are preserved in the simulation,
including the full details of its azimuthal correlations.

A reconstructed photon is considered ``matched'' to a prompt generator-level
(``truth'') photon when
they are separated by an angular distance
$\DR = \sqrt{ (\Delta \phi)^2 + (\Delta \eta)^2} < 0.2$.
If multiple reconstructed photons are within the matching window, only the highest-$\pT$
reconstructed photon is considered matched to the truth photon.

\section{Photon reconstruction}
\label{section:reco}
The electromagnetic shower associated with each photon, as well as the total
transverse energy in a cone surrounding it,
are reconstructed as described in Ref.~\cite{ATLAS:2011kuc}.
However, in a heavy-ion collision, it is important to
subtract the large UE from each event before the reconstruction
procedure is applied.
If it is not subtracted, photon transverse energies can be overestimated by
up to several~\GeV\ in the most central events,
and the isolation transverse energy
in a $\DR=0.3$ cone can be overestimated by about \mbox{60~\GeV}.
The procedure explained in Ref.~\cite{Aad:2012vca} is used to estimate
the energy density of the underlying event in each calorimeter cell.
It iteratively excludes jets from consideration in order to obtain the
average energy density in each calorimeter layer
in intervals of $\Delta\eta=0.1$, after accounting for
the elliptic
modulation relative to the event plane angle
measured in the FCal~\cite{ATLAS:2011ah,Poskanzer:1998yz}.
The algorithm provides the energy density as a function of $\eta$, $\phi$,
and calorimeter layer, which allows the event-by-event subtraction of
the UE in the electromagnetic and hadronic calorimeters.

After subtraction, the
residual deposited energies stem primarily from three sources:
jets, photons/electrons, and UE
fluctuations (including higher-order flow harmonics).
It should be noted that while this provides an estimate of the mean underlying
transverse energy as a function of $\eta$,
it is at present not possible to make further subtraction of more localized structures.

The \atlas\ photon reconstruction~\cite{ATLAS:2011kuc} is seeded by
clusters with $\ET > 2.5$~\GeV\ found using a sliding-window algorithm applied
to the second sampling layer of the electromagnetic calorimeter, which
typically contains over 50\% of the shower energy.
In the dense environment of the heavy-ion collision, the photon
conversion reconstruction
procedure is not performed, due to the large number of combinatoric pairs in more central collisions.
However, a substantial fraction of converted photons are still reconstructed
by the photon algorithm as, for high energy photon conversions,
the electron and positron are typically
close together when they reach the calorimeter,
while their tracks typically originate at
a radius too large to be well described by the tracking algorithm
that is used for heavy-ion collisions.
Thus, the photon sample analyzed here is a mix of converted and unconverted photons.
From simulations,
the overall conversion rate is found to be about 30\% in $|\eta|<1.37$ and
60\% in $1.52 \leq |\eta| < 2.37$.

The energy measurement is made using the three layers of the electromagnetic calorimeter and the presampler,
with a window size corresponding to $3 \times 5$ cells (in $\eta$ and $\phi$)
in the second layer in the barrel, and
$5 \times 5$ cells in the endcap region.
An energy calibration is applied to each shower to account for both its lateral leakage (outside the nominal window) and longitudinal leakage (into the hadronic calorimeter as well as dead material)~\cite{ATLAS:2011kuc}.
For converted photons, this window size can lead to an underestimate of
the photon candidate's energy, which is accounted for in the data analysis.
The transverse energy of the photon is defined as the calibrated cluster energy
multiplied by the sine of the polar angle
determined with respect to the measured event vertex.
The transverse momentum of the photon is identified with the measured transverse energy.

The fine-grained, longitudinally segmented calorimeter allows for a
detailed characterization of the shape of each photon shower, which can be
used to reject neutral hadrons while maintaining a high efficiency for photons.
Nine shower shape variables are used for each photon candidate:

The primary shape variables used can be broadly classified by which sampling layer is used.
The second sampling layer is used to measure
\begin{itemize}
\item $R_{\eta}$: the ratio of energies deposited in a $3\times7$ ($\eta \times \phi$) window to those deposited in a $7 \times 7$ set of cells in the second layer.
\item $R_{\phi}$: the ratio of energies deposited in a $3\times3$ ($\eta \times \phi$) window to those deposited in a $3 \times 7$ set of cells in the second layer.
\item $w_{\eta,2}$: the standard deviation in the $\eta$ projection of the energy distribution of the cluster in a $3\times 5$ set of cells in the second layer.
\end{itemize}
The hadronic calorimeter is used to measure the fraction of shower energy that is detected behind the electromagnetic calorimeter.  Only one of these is applied to each photon, depending on its pseudorapidity:
\begin{itemize}
\item $R_{\mathrm{\mathrm{had}}}$: the ratio of transverse energy measured in the hadronic calorimeter to the transverse energy of the photon candidate.
This quantity is used for $0.8\leq|\eta|<1.37$.
\item $R_{\mathrm{\mathrm{had1}}}$: the ratio of transverse energy measured in the first sampling layer of the hadronic calorimeter to the transverse energy of the photon candidate.
This quantity is used for photons with either $|\eta|<0.8$ or $|\eta| \geq 1.52$.
\end{itemize}
Finally, cuts are applied to five other quantities, measured in the
fine-granularity first layer, to reject neutral meson decays from jets.
In this finely segmented layer a search
for multiple maxima from electromagnetic decays of neutral hadrons is performed:
\begin{itemize}
\item $w_{\mathrm{s,tot}}$: the standard deviation of the energy distribution in the $\eta$ projection in the first sampling ``strip'' layer, in strip cell units.
\item $w_{\mathrm{s},3}$: the standard deviation of the energy distribution in three strips including and surrounding the cluster maximum in the strip layer, also in strip cell units.
\item $\Fside$: the fraction of energy in seven strips surrounding the cluster maximum, not contained in the three core strips.
\item $E_{\mathrm{ratio}}$: the asymmetry between the energies in the first and second maxima in the strip layer cells.  This quantity is equal to one when there is no second maximum.
\item $\Delta E$: the difference between the energy of the second maximum, and the minimum cell energy between the first two maxima.  This quantity is equal to zero when there is no second maximum.
\end{itemize}

In a previous ATLAS measurement~\cite{Aad:2010sp},
it was observed that the distributions of
the shower shape variables measured in data differ systematically
from those in the simulation.  To account for these differences, a set of correction factors
was derived, each of which changes the value of a simulated shower shape variable
such that its mean value matches that of the corresponding measured distribution.
For the measurements presented in this \ptype,
the same correction factors,
obtained by comparing $\pp$ simulations to the same quantities in data,
are used with no modification for the heavy-ion environment.
They were validated in the heavy-ion environment using electrons and positrons
from reconstructed $Z \rightarrow e^+ e^-$ decays from
the same LHC run.
It was observed that the magnitude and centrality dependence of the mean
values of the shape variables are well described
by simulations, within the limited size of the electron and positron sample.

\begin{figure*}[t!]
\begin{center}
\includegraphics[width=0.95\textwidth]{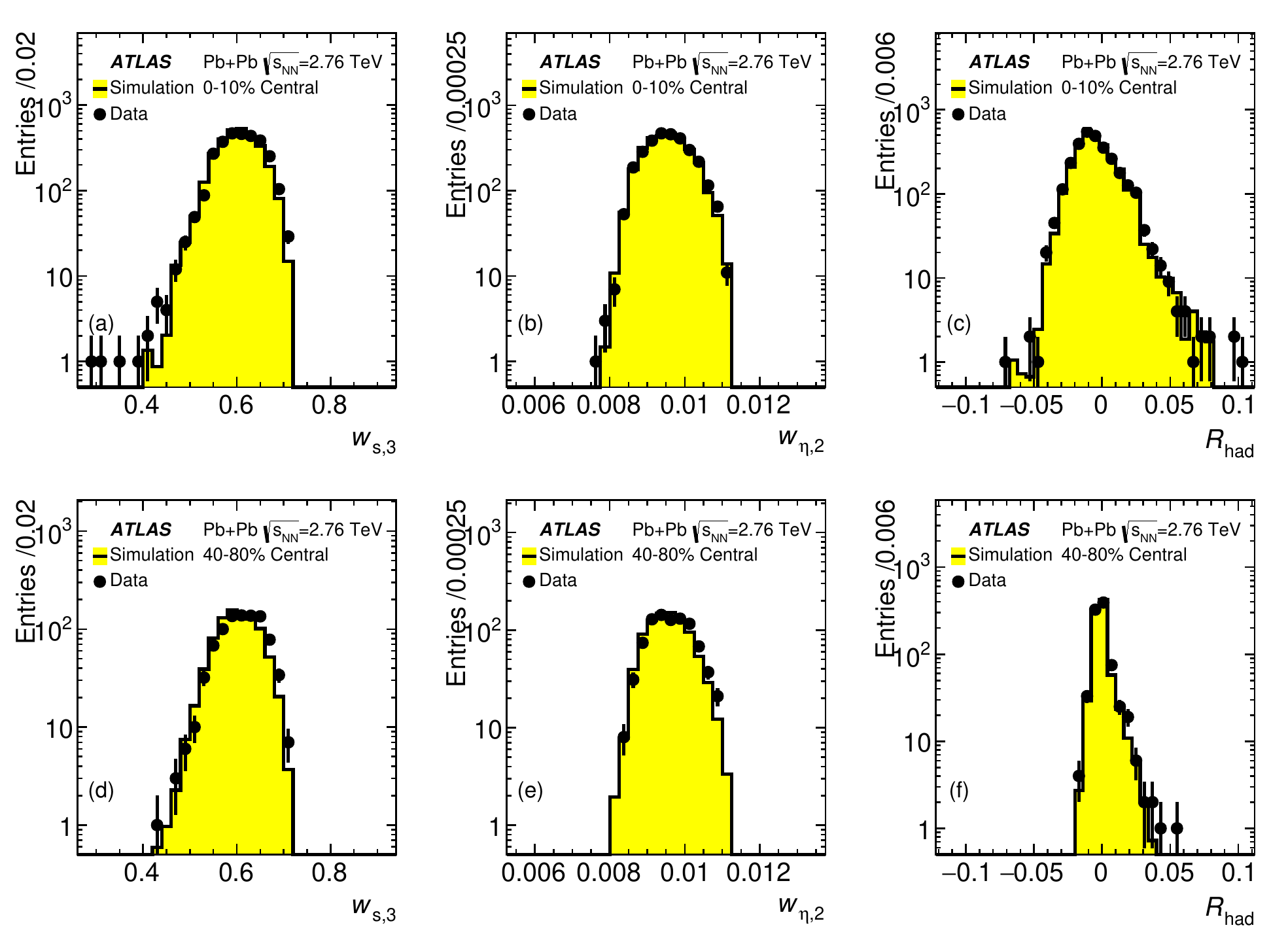}
\caption{ (Color online) Comparisons of distributions of three shower shape variables
  ($w_{\mathrm{s,3}}$, $w_{\eta,2}$ and $R_{\mathrm{had}}$) from data
  (black points) with simulation results after shower shape
  corrections (yellow histogram), for tight and isolated photons with
  reconstructed $35 \leq \pT < 44.1$~\GeV\ and $|\eta|<1.37$.  Events
  from the 0--10\% centrality interval are shown in the top row (a)-(c), while
  those from the 40--80\% interval are shown in the bottom row (d)-(f).
\label{figure:data_vs_simulation}
}
\end{center}
\end{figure*}

Figure~\ref{figure:data_vs_simulation} shows three typical distributions of shower
shape variables
for data from the 0--10\% and 40--80\% centrality intervals, each compared with
the corresponding quantities in the simulation.
The simulated distributions, after shower shape corrections,
are all normalized to the number of counts
in the corresponding data histogram.
The data contain some admixture of neutral hadrons, so complete agreement should not be
expected in the full distributions.
The admixture of converted photons, which depends on the amount of material in front of the
electromagnetic calorimeter, and thus the pseudorapidity of the photon, is not accounted for
in the analysis, but there is
good agreement of the shower shape variable distributions
between data and simulation.  Converted photons tend to
have wider showers than unconverted photons,
and so substantially broaden the shower shape variables.

The electromagnetic-energy-trigger efficiency was investigated using a sample of
minimum-bias data, where the primary triggers did not select on particular high-$\pT$ activity.
Using these, the probability for photon candidates selected for this analysis to match
a first-level trigger
with \mbox{$E_{\mathrm {T,trig}}> 16$} \GeV\ and $\Delta R <0.15$
exceeds 99\% for well-reconstructed photon candidates
with $\pT \geq 22$ GeV and over the full centrality range.
In the more central events, the underlying-event contribution to the photon candidate
reduces the effective threshold down by several GeV relative to the more peripheral events.
To work in the plateau region, the minimum \pT\ required in this analysis is 22 GeV.

Photons are selected for offline analysis using a variation of the ``tight'' selection criteria developed for the
photon analysis in \pp\ collisions~\cite{Aad:2010sp}, necessitated by the additional fluctuations
in the shower shape variables induced by the underlying event in heavy-ion collisions.
Specific intervals are defined for all nine shower
shape variables, and are implemented in a $\pT$-independent, but $\eta$-dependent
scheme.
The intervals for each variable are defined
to contain 97\% of the distribution of isolated
reconstructed photons matched to isolated truth photons
with a reconstructed $\pT$ in the region
\mbox{$40 \leq \pT < 60$~$\GeV$} in the 0--10\% centrality interval (where the UE fluctuations are largest), using
the isolation criteria described in the next section.

In order to derive a data-driven estimate of the background candidates
from jets, a ``nontight'' selection criterion is defined,
which is particularly sensitive to neutral hadron decays.
For this selection, a photon candidate
is required to fail at least one of four shower shape selections in the first calorimeter
layer: $w_{\mathrm{s,3}}$, $\Fside$, $E_{\mathrm{ratio}}$ and $\Delta E$.
These reversed selections enhance the probability of accepting neutral hadron decays from jets, via
candidates with a clear double shower structure (via $E_{\mathrm{ratio}}$ and $\Delta E$)
as well as candidates in which the two showers may have merged
(via $w_{\mathrm{s,3}}$ and $\Fside$)~\cite{Aad:2010sp}.

While the photon energy calibration is the same as used for $pp$ collisions,
based in part on measurements of
$Z$ bosons decaying into an electron and a positron, and validated with $Z\rightarrow \ell\ell+\gamma$
events~\cite{PERF-2010-04}, the admixture of converted and unconverted
photons leads on average to a small underestimate of the photon energy
in \PbPb\ events, since the energies of converted photon clusters is
typically reconstructed in a larger region in the calorimeter.
This is quantified in the simulation by the
mean fractional difference between the reconstructed and the truth
photon transverse momenta
$(\pT^{\mathrm{reco}} - \pTtruth) / \pTtruth \equiv
\Delta \pT / \pTtruth$, obtained from simulation.
For matched photons, the average deviation from the truth photon \pT\
is the largest at low photon $\pT$ and is typically within $1\%$ for $\pT>44$ GeV.
The fractional energy resolution, determined
by calculating the standard deviation of the same quantity
in smaller intervals in $\pTtruth$,
ranges from 4.5\% for
$22 \leq \pTtruth < 26$~\GeV\ to 1.5\% for $\pTtruth=200$~\GeV\ for $|\eta|<1.37$
and from 6\% to 3\% for $1.52 \leq |\eta| < 2.37$.
The effects of energy scale and resolution
are corrected for by using bin-by-bin correction factors described below.

The isolation transverse energy $\ETiso$
is the sum of transverse energies in calorimeter cells
(including hadronic and electromagnetic sections) in a
cone of size $\Riso$ around the photon direction.
The photon energy is removed by excluding a
central core of cells in a region corresponding to 5$\times$7 cells in the second
layer of the EM calorimeter.
The cone size is chosen to be $\Riso = 0.3$,
to reduce the sensitivity to UE fluctuations.
The isolation criterion is
$\ETiso < 6$~$\GeV$.
An additional correction, based on simulations and
parameterized primarily by the photon energy and $\eta$,
is then applied to the calculated isolation transverse energy to minimize the effects of
photon shower leakage into the isolation cone.
It typically amounts to a few percent of the
reconstructed photon transverse energy.

\begin{figure}[t!]
\begin{center}
\includegraphics[width=0.5\columnwidth]{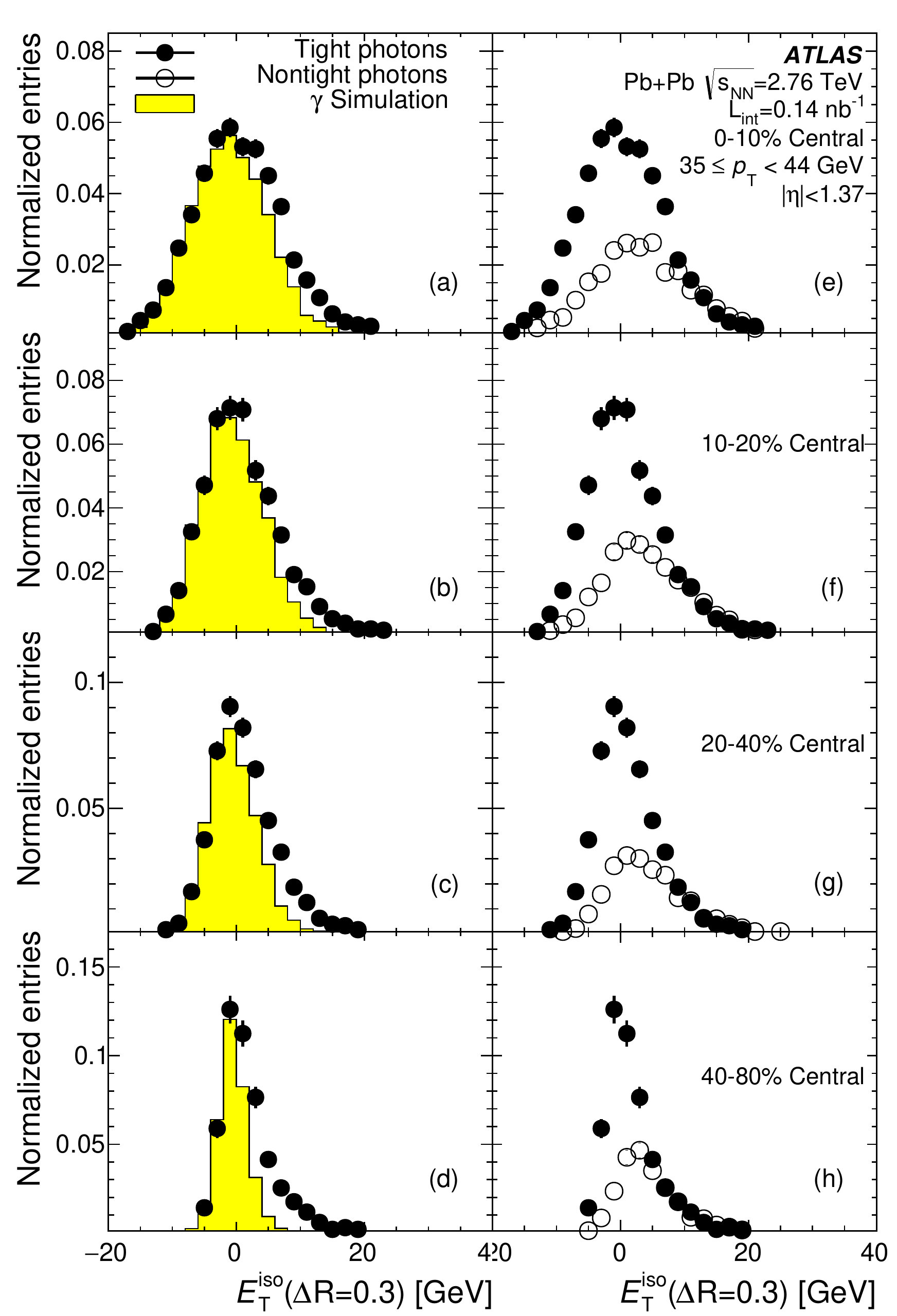}
\caption{  (Color online) Distributions of photon isolation transverse energy in a
  $\Riso = 0.3$ cone for the four centrality bins in data (black
  points, normalized by the number of events and by the histogram bin
  width), for photons with $35 \leq \pT < 44.1$~$\GeV$.  In the left
  column (a)-(d) simulations (yellow histogram) are normalized to the data so
  that the integrals in the range $\ETiso<0$ are the same.  The
  corresponding sample of nontight photon candidates, normalized to
  the distribution of tight photons for $\ETiso \geq 8$~\GeV\ is shown
  overlaid on the tight photon data in the right column (e)-(h) to illustrate
  the source of the photons with large $\ETiso$.  }
\label{figure:comp_etcone}
\end{center}
\end{figure}

The left column of
Fig.~\ref{figure:comp_etcone} shows the distributions of $\ETiso$ for
tight photon candidates with $35 \leq \pT < 44.1$~\GeV\
as a function of collision centrality, compared with simulated distributions.
The data and simulations are normalized so the integrals of $\ETiso<0$, where no significant
background from jet events is expected, are the same.
Both, the simulated and the measured $\ETiso$ distributions grow noticeably wider with increasing centrality;
as the UE subtraction only accounts for the mean energy in an $\eta$ interval, local fluctuations are still present.
Furthermore, in the data, an enhancement in events with $\ETiso>0$
is expected from the jet background.
The $\ETiso$ distribution for a sample enhanced in backgrounds is shown
in the right column of Fig.~\ref{figure:comp_etcone}, which shows the isolation distribution for the nontight candidates in the same $\pT$ interval.  For
larger values of $\ETiso$, the distributions from the tight and nontight samples have similar shapes.
The distributions are normalized to the integral of the tight photon candidate distribution
in the region $\ETiso >8$~$\GeV$.

After applying the tight selection and an isolation
criterion of $\ETiso < 6$~\GeV\ to the 0--80\% centrality sample, there
are 62\,130 candidates with $\pT \geq 22.0$~\GeV\ within
$|\eta|<1.37$ and 30\,568 candidates within $1.52 \leq |\eta| < 2.37$.

\section{Yield extraction}

\label{section:bins}
The kinematic intervals used in this analysis are defined as follows.  For each centrality interval,
as described in Sect.~\ref{sect:collision},
the photon kinematic phase space is divided
into intervals in photon $\eta$ and $\pT$.
The two primary regions in $\eta$ are $|\eta|<1.37$ (``central $\eta$''),
$1.52 \leq |\eta| < 2.37$ (``forward $\eta$'').
The $\pT$ intervals used are logarithmic, and are
$17.5 \leq \pT < 22$ ~\GeV\ (only used in simulations),
$22.0 \leq \pT < 27.8$~\GeV,
$27.8 \leq \pT < 35.0$~\GeV,
$35.0 \leq \pT < 44.1$~\GeV,
$44.1 \leq \pT < 55.6$~\GeV,
$55.6 \leq \pT < 70.0$~\GeV,
$70.0 \leq \pT < 88.2$~\GeV,
$88.2 \leq \pT < 140$~\GeV,
and
$140 \leq \pT < 280$~\GeV .

Prompt photons are defined as photons produced in the simulation of the hard process,
either directly or radiated from a primary parton, via
a truth particle-level isolation transverse energy selection of
$\ETiso < 6$~\GeV.
The truth-level $\ETiso$ is defined using all final-state particles except
for muons and neutrinos in a cone of $\DR=0.3$ around the photon direction.
To account for the underlying event in the hard process, the mean energy
density is estimated for each simulated event using
the jet-area method described in Ref.~\cite{Aad:2010sp}.

For each interval in $\pT$, $\eta$ and centrality ($\Cent$), the per-event
yield of photons
is defined as
\begin{equation}
\label{eq:yield}
\frac{1}{\Nevt(\Cent)} \frac{ dN_{\gamma}}{d\pT} ( \pT, \eta, \Cent)  =
\frac{ \NsigA \Unf(\pT, \eta, \Cent) \Cont(\pT, \eta, \Cent) }{\Nevt(\Cent) \efftot(\pT, \eta, \Cent) \Delta\pT},
\end{equation}
where $\NsigA$ is the background-subtracted yield,
$\Unf$ is a factor that corrects for the bin migration due to the photon energy resolution and
any residual bias in the photon-energy scale,
$\Cont$ is a factor that corrects for electron contamination from \W\ and \Z\ bosons,
$\efftot$ is the combined photon reconstruction and identification efficiency,
$N_{\mathrm{evt}}$ is the number of minimum-bias
events in centrality interval $\Cent$,
and $\Delta \pT$ is the width of the
transverse momentum interval.

\begin{figure}[t!]
\begin{center}
\includegraphics[width=0.35\textheight]{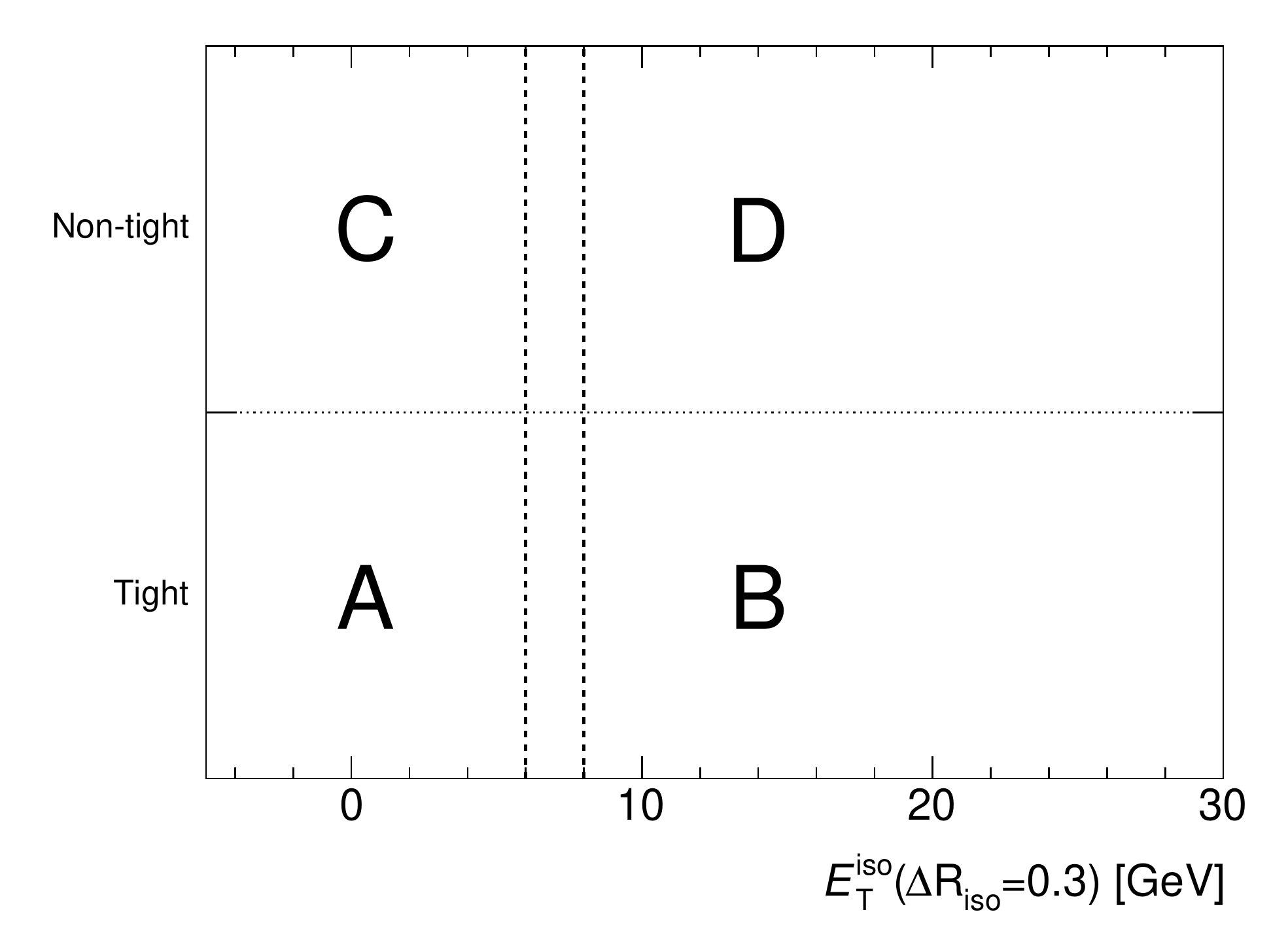}
\caption{
\label{figure:doublesideband}
 (Color online) Illustration of the double sideband approach, showing the two axes for
partitioning photon candidates: region A is the ``signal region''
(tight and isolated photons), region B contains tight, nonisolated
photons, region C contains nontight isolated photons, and region D
contains nontight and nonisolated photons.  }
\end{center}
\end{figure}

The technique used to subtract the background from jets from the measured yield of photon candidates
is the ``double sideband'' method, used in
Refs.~\cite{Aad:2010sp,Aad:2011tw,Aad:2013zba}.
In this method, photon candidates are partitioned along
two dimensions, illustrated in Fig.~\ref{figure:doublesideband}.
The four regions are labeled A, B, C and D and correspond to the four categories expected for reconstructed photons
and background candidates:
\begin{itemize}
\item {A: tight, isolated photons:} Signal region for prompt, isolated photons.
\item {B: tight, nonisolated photons:} A region expected to contain nonisolated photons produced in the vicinity of a jet or an upward
UE fluctuation, as well as hadrons from jets with shower shapes similar to those of a tight photon.
\item {C: nontight, isolated photons:}
A region containing isolated neutral hadron decays, e.g. from hard-fragmenting jets,
as well as real photons that have a shower-shape fluctuation that fails the tight selection.
\item {D: nontight, nonisolated photons:}
A region populated by neutral hadron decays
within jets, but which have both a small admixture of photons that fail the tight selection
and are accompanied by a local upward fluctuation of the UE.
\end{itemize}

The nontight and nonisolated photons are used to estimate the
background from jet events in the signal region A.  This is appropriate
provided there is no correlation between the axes for
background photon candidates, e.g. that the probability of a neutral hadron decay
satisfying
the tight or nontight selection criteria is not dependent on whether or not it is isolated.
This was studied using a sample of high-$\pT$ photon candidates from the large sample of \PYTHIA\ inclusive jet events.
Possible correlations, parameterized by the $\Rbkg$ ratio~\cite{Aad:2010sp},
$\Rbkg = N_{\mathrm A}^{\mathrm{bkg}} N_{\mathrm D}^{\mathrm{bkg}} / ( N_{\mathrm B}^{\mathrm{bkg}} N_{\mathrm C}^{\mathrm{bkg}})$,
are taken as a systematic uncertainty, as discussed in
Sect.~\ref{section:uncertainties}.

If there is no leakage of signal from region A to the other nonsignal regions (B, C and D), the
double sideband approach utilizes the ratio of counts in C to D to extrapolate the measured number of counts in region B to
correct the measured number of counts in region A, i.e.
\begin{equation}
\label{eq:doublesideband}
N^{\mathrm{sig}} = \NsigA = \NobsA - \NobsB \frac{\NobsC}{\NobsD} .
\end{equation}
Leakage of signal into the background regions needs to be removed before attempting to
extrapolate into the signal region.
A set of ``leakage factors'' $\ci$ are calculated to extrapolate the number of signal events in region A into the other
regions.
The leakage factors are calculated using the \PYTHIA\ simulations
in intervals of reconstructed photon $\pT$
as $\ci = \Nsigi / \NsigA$, where $\NsigA$ is the number of simulated tight, isolated photons.
In the 40--80\% centrality interval, for $|\eta|<1.37$
and for $22 \leq \pT < 280$~\GeV, $\cB$ is generally less than 0.01, $\cC$ ranges from
0.09 to 0.02, and $\cD$ is less than 0.003.
In the 0--10\% centrality interval and over the same $\pT$ range, $\cB$ ranges from 0.08 to 0.11, $\cC$ ranges
from 0.13 to 0.04, and $\cD$ is O(1\%) or less.  Except for $\cB$, which reflects the different isolation
distributions in peripheral and central events, the leakage factors are of similar
magnitude to those derived in the $\pp$ data analysis~\cite{Aad:2010sp}.

Including these factors, and the correlation parameter $\Rbkg$, the formula
becomes
\begin{equation}
\label{eq:ds}
\NsigA = \NobsA - \\
\Rbkg \left( \NobsB-\cB \NsigA \right) \frac{ \left( \NobsC - \cC \NsigA \right) }{ \left( \NobsD - \cD \NsigA \right) } .
\end{equation}
Equation~(\ref{eq:ds}) is solved for the yield of signal photons $\NsigA$, with $\Rbkg$ assumed to be 1.0.
The statistical uncertainties in the number of signal photons for each centrality, $\eta$ and $\pT$
interval are evaluated with 5\,000 pseudo-experiments.
For each pseudo-experiment,
the parameters $\NobsA$, $\NobsB$, $\NobsC$ and $\NobsD$ are sampled
from a multinomial distribution with the probabilities given by the
observed values divided by their sum.
The values of $\NsigA$, $\NsigB$, $\NsigC$, and $\NsigD$ used to
determine the leakage factors in each experiment, are themselves sampled
from a Gaussian distribution with the parameters determined by
the means of the simulated distributions and their statistical
uncertainties.
Pseudo-experiments where the leakage correction is negative are discarded,
to exclude trials where the extracted yield is larger than $\NobsA$.
The standard deviation of the distribution of $\NsigA$ obtained from the set of
pseudo-experiments is taken as the statistical uncertainty.

\begin{figure*}[t!]
\begin{center}
\includegraphics[width=0.95\textwidth]{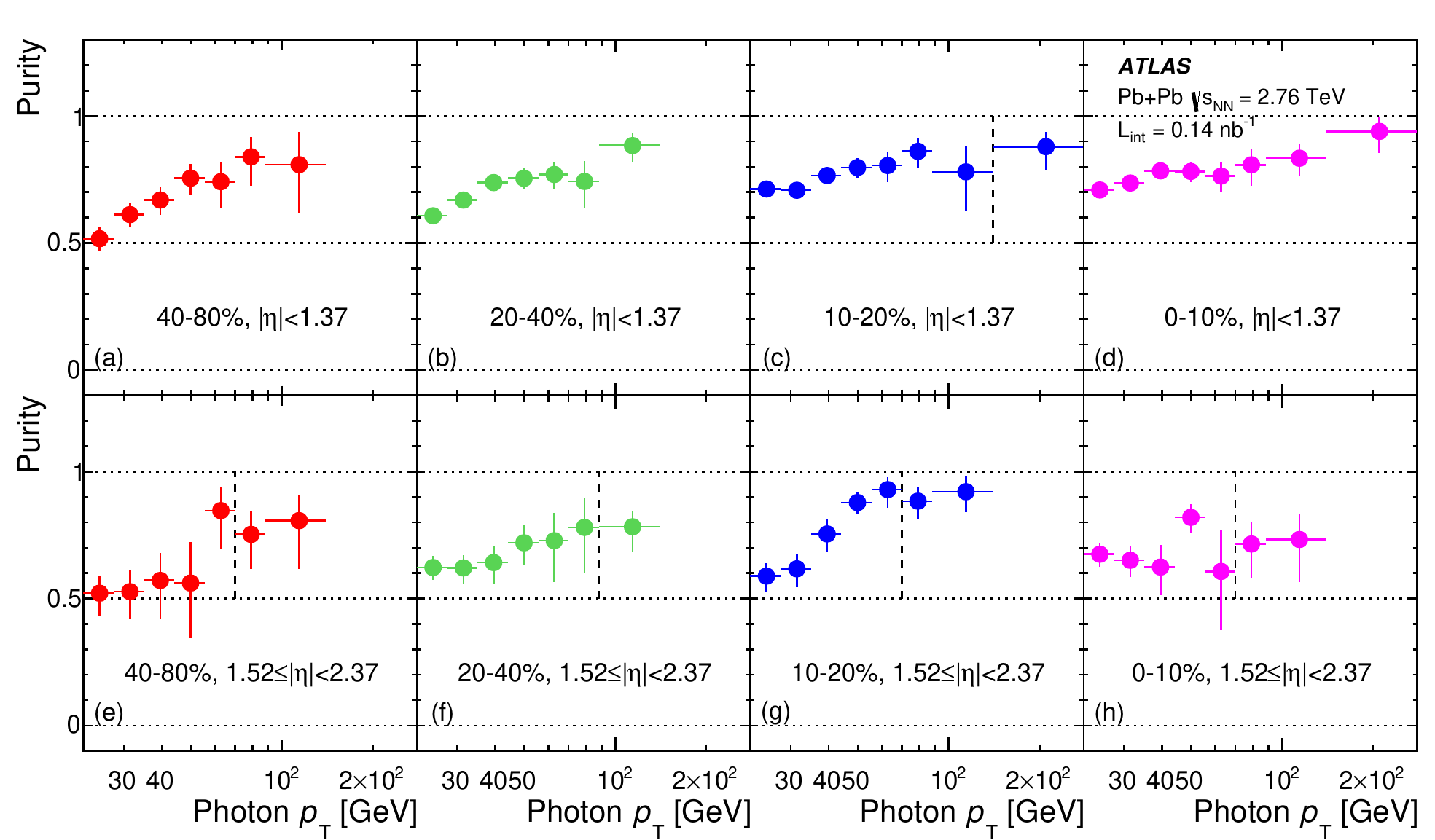}
\caption{ (Color online)  Photon purity as a function of collision centrality (left to
  right) and photon $\pT$, for photons measured in $|\eta|<1.37$ ((a)-(d))
  and $1.52 \leq |\eta|<2.37$ ((e)-(h)).  The $\pT$ intervals
  to the right of the vertical dotted line indicated in some bins use
  the extrapolation method described in the text to account for low
  event counts in the sidebands.
\label{figure:purity}
}
\end{center}
\end{figure*}

The purity of the photon sample in the double sideband method is then defined as
$P=\NsigA/\NobsA$.
The extracted values of $P$ are shown in Fig.~\ref{figure:purity}
as a function of transverse momentum in the four measured centrality intervals and two $\eta$ intervals.
In all four centrality and both $\eta$ intervals, the purity increases from about 0.5 at the lowest
$\pT$ interval to 0.9 at the highest $\pT$ intervals, with typically lower values in the
forward $\eta$ region.
The statistical uncertainty in the purity is determined specifically using the pseudo-experiments described above, and by
using the boundaries defined by the highest and lowest 16\% of the purity distributions to determine
the upper and lower asymmetric error bars.

For kinematic regions in which the number of candidates in the sidebands are small, particularly
at the highest $\pT$ values, the population of those sidebands are re-estimated using a
data-driven approach.
For this, the ratio of each sideband (B, C, and D) to region A as a function of $\pT$ is measured and extrapolated
linearly in $1/\pT$, utilizing all of the available data up to $\pT = 140$ GeV.
It should be noted that the purity merely represents the outcome of the
sideband subtraction procedure, and is not used as an independent correction factor.
The several points for which this extrapolation is utilized are those to the right of the vertical dotted line in
several of the Fig.~\ref{figure:purity} centrality intervals.

\begin{figure*}[t!]
\begin{center}
\includegraphics[width=0.95\textwidth]{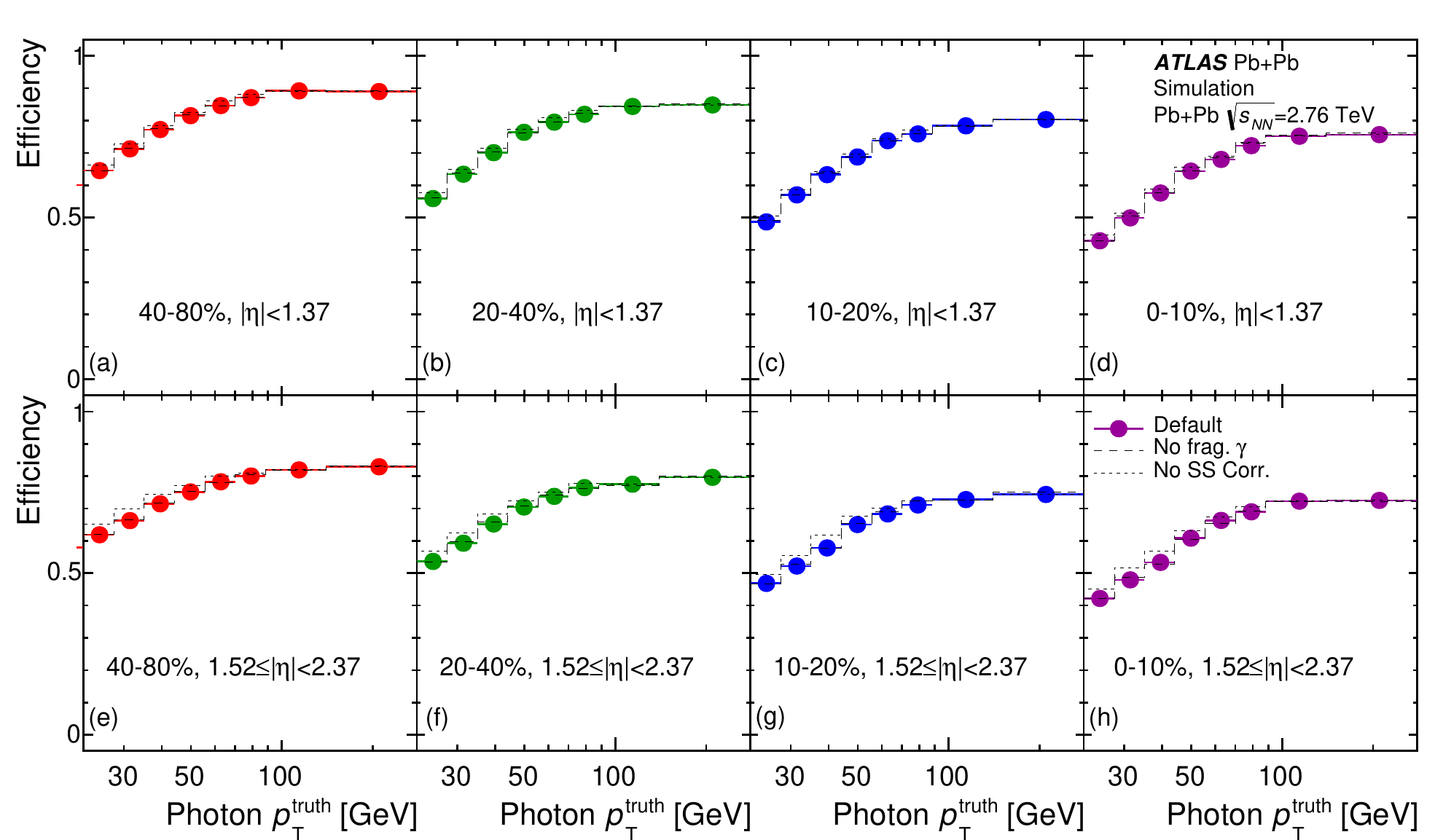}
\caption{
\label{figure:efficiency}
 (Color online) Total photon efficiency as a function of photon $\pT$ and event
centrality averaged over $|\eta|<1.37$ ((a)-(d)) and $1.52 \leq
|\eta|<2.37$ ((e)-(h)).  Variations of the efficiency from removing
the small corrections to the simulated shower-shape variable, and from
removing fragmentation photons from the simulations are shown by
dotted and dashed lines, respectively.  }
\end{center}
\end{figure*}

The reconstruction efficiency is the fraction of tight, isolated photons
matched to the truth photons defined above ($\ETiso<6$~\GeV), according to the criterion
specified in Sect.~\ref{section:sim}.
The true photon $\pT$ is used in the numerator and the denominator, while the reconstructed $\eta$ is used in the numerator
to estimate the very small inflow and outflow of photons in the large $\eta$ intervals used in the analysis.
The total efficiency can be factorized into the product of three contributions:
\begin{itemize}
\item {Reconstruction efficiency:} the probability that a photon is reconstructed with a
$\pT$ greater than 10~\GeV.  In the reconstruction algorithm, the losses primarily stem from a subset of photon conversions, for which the photon is reconstructed as an electron (``photon to electron leakage'').
The losses are typically 5\% near $\eta=0$ and increase to about 10\% at forward angles,
and are found to be approximately constant as a function of transverse momentum and centrality.
\item {Identification efficiency:} the probability that a reconstructed photon passes the tight identification selection criteria.
\item {Isolation efficiency:} the probability that a photon that
would be reconstructed and pass the identification selection criteria, also passes the chosen isolation selection.  The large fluctuations from the UE
in heavy-ion collisions can lead to a photon being found in the nonisolated region.
\end{itemize}
Figure~\ref{figure:efficiency} shows the total efficiency for each centrality and $\eta$
interval as a function of photon $\pT$.
The primary systematic uncertainties on the efficiency were evaluated by removing the
small correction factors applied to the simulated shower shapes, and by excluding fragmentation photons from
the sample used to derive the efficiencies.  The contribution from each individual shower-shape selection
is small, and so the effect on the efficiency is typically small,
but the cumulative effect is as large as 10\% in the lowest $\pT$ intervals in the forward $\eta$ region.
Similar correction factors were calculated using the \SHERPA\
simulations, and they are found to be consistent
with the \PYTHIA\ calculations in all considered centrality and $\eta$ regions.

To account for the residual deviations of the measured photon $\pT$ from the true $\pT$,
stemming primarily from converted photons treated as unconverted,
and from the photon energy resolution, the data are corrected using
a bin-by-bin correction technique~\cite{Aad:2010sp}
to generate the correction factors $\Unf$.
For each interval in centrality and $\eta$, a response matrix is
formed by correlating the reconstructed $\pT$ with the truth $\pT$ for truth-matched photons.
The projections onto each $\pT$ interval along the truth axis $T_{\mathrm{i}}$
and the reconstructed axis $R_\mathrm{i}$ are then constructed for
each centrality and $\eta$ interval and their ratio
$C_\mathrm{i}=T_\mathrm{i}/R_\mathrm{i}$
is formed to calculate the correction
in the corresponding $\pT$ interval.
To reduce the effect of statistical fluctuations, the $C_{\mathrm{i}}$ values were
fit to a smooth functional form before applying to the data, with the deviations of the extracted correction
factors from the fit being generally O(1\%).
In the lowest $\pT$ interval ($22.1 \leq \pT < 28$~\GeV),
the correction factors deviate from unity by
$+(6 \ndash 9)\%$ in the central $\eta$ region
and $+(8 \ndash 13)\%$ in the forward $\eta$ region
(the first number for the 40--80\% centrality interval, and the second
for the 0--10\% interval).
They approach unity rapidly as a function of $\pT$ and
in the highest $\pT$ interval are
$-2\%$ in the central $\eta$ region and $+2\%$ in the forward $\eta$ region.
The reconstructed spectral shapes were compared between
simulation and data, and
were found to agree within statistical uncertainties.
Thus, no reweighting of the simulated spectrum
was performed before calculating the bin-by-bin factors.

Samples of simulated $W$ and $Z$ bosons decaying to
electrons or positrons, based on $\POWHEG$~\cite{Frixione:2007vw}
interfaced to the $\PYTHIA 8$ generator (version 8.175)~\cite{Sjostrand:2007gs},
were used to study the estimated
contamination rate relative to the total photon rates expected from $\jetphox$.
The raw contamination electron rates were corrected using the photon total efficiency.
The difference in the extracted cross section of contamination electrons between the
most peripheral and the most central events was found to be modest.
Therefore the centrality dependence is neglected and the
cross sections calculated for the most central events are used in all centrality intervals.
Based on this study, it was estimated that the largest
background of the $W$ and $Z$ background
is expected in the $35 \leq \pT < 44.1$~\GeV\ interval with a magnitude of about 8\% in the
forward pseudorapidity region, and about 5\% in the central region.
In other bins the correction is smaller, and in most bins it is less than $2\%$
in the central $\eta$ region and less than $3\%$ in the forward $\eta$ region.

\section{Systematic uncertainties}
\label{section:uncertainties}

\begin{table*}
\caption{
\label{yield_syst}
Relative systematic uncertainties, expressed as a percentage, on the
efficiency-corrected yields for selected $\pT$ and centrality
intervals in the two $\eta$ intervals.  } \footnotesize
\begin{center}
\begin{tabular}{|p{2.5cm}|c|c|c|c|c|c|c|c|}
\hline \hline
$\eta$ & \multicolumn{4}{c|}{$|\eta|<1.37$} & \multicolumn{4}{c|}{$1.52 \leq |\eta|<2.37$} \\
\hline
Centrality & \multicolumn{2}{c|}{40--80\%} & \multicolumn{2}{c|}{0--10\%} &\multicolumn{2}{c|}{40--80\%} & \multicolumn{2}{c|}{0--10\%}  \\
\hline
$\pT$ [GeV] & 22--28 & 55.6--70  & 22--28  & 70--88.2  & 22--28  & 55.6--70  & 22--28  & 70--88.2  \\
\hline
\hline
$\gamma \rightarrow e$ leakage& 1& 1& 1& 1& 1& 1& 1& 1 \\
\hline
\mbox{Shower shape corr.}& 3& 2& 5& 3& 6& 2& 9& 3 \\
\hline
Isolation& 7& 5& 6& 8& 6& 10& 5& 9 \\
\hline
Frag. photons& $<1$& $<1$& 1& 2& 1& $<1$& 2& 2 \\
\hline
\hline
Leakage factors& 10& 4& 12& 9& 7& 1& 15& 10 \\
\hline
\mbox{Nontight criteria}& 4& 4& 3& 3& 7& 6& 6& 5 \\
\hline
$\Rbkg$ & 21& 7& 13& 6& 20& 4& 15& 11 \\
\hline
\hline
Energy scale& 5& 5& 5& 5& 5& 5& 5& 5 \\
\hline
\mbox{$W/Z$ contamination}& $<1$& 1& $<1$& 1& 1& 1& 1& $1$ \\
\hline
Cent. weight& 4& 1& 1& $<1$& 3& 1& 4& $<1$ \\
\hline
$\eta$ leakage& $<1$& $<1$& $<1$& $<1$& 2& 1& 2& 2 \\
\hline
Total [\%]& 26& 12& 21& 15& 25& 14& 25& 19 \\
\hline

\hline \hline
\end{tabular}
\end{center}
\end{table*}

\begin{table*}
\caption{
\label{ratio_syst}
Relative systematic uncertainties, expressed as a percentage, on the
ratio of the yields in the forward $\eta$ region and those in the
central $\eta$ region $\RFC$ for selected $\pT$ and centrality
intervals in the two $\eta$ intervals.  } \footnotesize
\begin{center}
\begin{tabular}{|p{2.5cm}||c|c|c|c|}
\hline
\hline
Centrality & \multicolumn{2}{c|}{40--80\%} & \multicolumn{2}{c|}{0--10\%} \\
\hline
$\pT$ [GeV] & 22--28 & 55.6--70  & 22--28  & 70--88.2  \\
\hline
\hline
$\gamma \rightarrow e$ leakage & 1 & 1 & 1 & 1 \\
\hline
\mbox{Shower shape corr.} & 3 & 0 & 4 & 0 \\
\hline
Isolation & 9 & 9 & 4 & 4 \\
\hline
Frag. photons & 1 & 0 & 1 & 1 \\
\hline
\hline
\mbox{Nontight criteria}& 3 & 3 & 4 & 4 \\
\hline
Leakage factors & 2 & 2 & 2 & 4 \\
\hline
$\Rbkg$ & 1 & 4 & 2 & 6 \\
\hline
\hline
Energy scale & 7 & 7 & 7 & 7 \\
\hline
\mbox{$W/Z$ contamination} & 0 & 1 & 0 & 1 \\
\hline
Cent. weight & 1 & 0 & 4 & 1 \\
\hline
$\eta$ leakage & 2 & 1 & 2 & 2 \\
\hline
Total [\%] & 13 & 13 & 11 & 11 \\
\hline

\hline
\hline
\end{tabular}
\end{center}
\end{table*}

The following systematic uncertainties are accounted for in this analysis.
They are broadly classified into uncertainties that affect the efficiency, those that affect
the yield extraction, and several other additional effects.

The systematic uncertainties that primarily affect the total efficiency are:
\begin{itemize}
\item { Photon-to-electron leakage:}
The misidentification of photons as electrons, due to conversions, was
studied using a sample simulated
with extra material, and is found to be less than a 1\% effect
on the reconstruction efficiency, since these photons are considered unrecoverable.
\item{ Shower-shape corrections:}
To assess the cumulative effect of the small shower-shape corrections
applied to mitigate the differences between data and simulation, the corrections are removed and the difference in the recalculated yields taken as a conservative
systematic uncertainty.  This is a smaller effect at higher $\pT$ but is as
large as $9\%$ at low $\pT$ in the forward $\eta$ region.
\item { Isolation criteria:}
To assess the impact of differences between the underlying $\ETiso$ distributions
in data and simulation, several changes in the isolation selection were made.
In one case, the cone size was changed to $\Riso=0.4$ and the $\ETiso$ selection
enlarged to \mbox{10~\GeV}. In the second, the $\ETiso$ selection
was varied up and down by 2~\GeV.
Finally, the gap along the $\ETiso$ axis between regions A/C and B/D was removed.
In all of these cases, the selections were similarly adjusted in simulation.
In general, the variations in the yields show only a weak dependence on $\pT$.
To reduce the effect of statistical fluctuations, the variations as
a function fo $\pT$ are fit to constants
over $22 \leq \pT < 44.1$~\GeV\ and $44.1 \leq \pT < 140$~\GeV, and the most signficant variation is
applied symmetrically to all points in that $\pT$ region.  If the fit value
is consistent with zero, then the variation is reduced by half to avoid overcounting
the statistical fluctuations.
For the
foward-central ratios,
the varations are fit with a single function over $22 \leq \pT < 70$~$\GeV$.
In several cases, changing the isolation selection led to O(10\%) changes
that were
clearly consistent with statistical fluctuations.  In these cases, the variation was reduced to be 5\%, similar to the adjacent centrality interval.

The shower leakage corrections were been varied by 1\% of the
measured photon $\pT$ in data, but not in simulation, to account for possible defects in
the correction.

\item { Fragmentation contribution:}
Excluding the fragmentation photons from the
simulation sample has typically less than a 2\% effect on the
final yields over the full $\pT$ range.
\end{itemize}

The systematic uncertainties that primarily affect the purity of the photon sample in each kinematic and centrality interval are:
\begin{itemize}
\item{ Leakage factors:}
To test the sensitivity to mismodeling of the shower fluctuations that lead to leakage into sideband regions C and D,
the leakage factors were conservatively varied up and down by 50\%.  The magnitude is given by
the difference between the leakage factors in the 40--80\% peripheral events, where the underlying event does not
cause large extra fluctuations, and the 0--10\% most central events.
This leads to up to 10\% variations at low $\pT$, while the effect at higher $\pT$ is below 5\%.
\item{ Nontight definition:}
In order to assess the sensitivity to the choice of nontight criteria, which allow background into the analysis, the nontight definition was changed from four reversed conditions, to five (adding $w_\mathrm{s,tot}$) and two (using just $\Fside$ and $w_{\mathrm{s,3}}$).
Similar to isolation criteria variations, fits to constant values in two $\pT$
intervals (and one interval for the forward-central ratios) were performed to smoothen
the bin-to-bin statistical fluctuations.
In the central $\eta$ interval, the variation is typically less than 5\%
while it is 7\% or less in the forward $\eta$ interval.
\item{ Correlation of tight and isolation axes:}
The large inclusive-jet $\PYTHIA$ samples were used to study possible correlations between the
tight selection criteria and the isolation transverse energy.
This is characterized by calculating $\Rbkg$
for the backgrounds from jets, where the candidate is not matched to a truth
photon.
After integrating over centrality and $\pT$,
$\Rbkg$ was found to vary by about 10\% in the central $\eta$ region and 20\% in the forward $\eta$ region, albeit with large statistical uncertainties.
A conservative variation of $\pm20\%$ was propagated through the analysis, which gives up to a 20\% change at
low $\pT$ where the purity is lowest, decreasing to typically less than 10\% at higher $\pT$.
\end{itemize}

Uncertainties that pertain to corrections on the energy scale, electron contamination and centrality are described here.
\begin{itemize}
\item { Energy scale and resolution corrections:}
The effect of the energy scale and resolution
from variations in material, different energy calibration schemes, and known differences between data and simulations
in $\pp$ collisions are propagated into the bin-by-bin correction factors.
The overall variation from the known sources is typically
found to be below 2--3\%, and is approximately constant in $\pT$, but grows at high $\pT$ in the forward $\eta$
region.
However, the extra-material sample shows a small, but systematic, overall shift in the reconstructed energy
scale which is approximately independent of $\pT$ and centrality, but is larger in the forward $\eta$ interval.
Based on this, an overall uncertainty of $5\%$ is assigned in the central $\eta$ region and
in the forward region, except in the forward region above 88.2~\GeV, where 7\% is assigned.  In the ratio, these errors are treated as fully uncorrelated between the two $\eta$ regions.

\item { Electron contamination:} The contamination from $W$ and $Z$ bosons was estimated to be largest in
the two $\pT$ intervals between 35 and 55.6~\GeV, and smaller in the
other $\pT$ intervals.  Since the calculation does not
account for the different expected leakage of the electrons into
the different sidebands, and since the number of
$Z$ bosons in the heavy-ion data is too low to determine this fully, 50\% of the contamination has
been assigned as an uncertainty, leading to a maximum of 4\% in one $\pT$ interval in
the forward region and smaller in all other intervals.

\item { Centrality:}
The uncertainty on $\meanTAA$ for each centrality interval
is given in Table~\ref{table:centrality} and is shared by
all $\pT$ and $\eta$ intervals for that centrality interval.
In addition, the
effect of reweighting the simulated FCal distribution generally
has a less than 2\%  effect on the final yields,
although the impact can increase to up to 4\% at low $\pT$
in the forward $\eta$ interval.

\item { $\eta$ leakage:}
To address the effect of photons migrating in and out of the large $\eta$ intervals when calculating the efficiency,
the true $\eta$ was also used for the efficiency calculations and was found to have a 1--2\% overall effect,
reaching the larger end of this range in the forward $\eta$ region.
\end{itemize}

For the absolute yields, all contributions are added in quadrature.
For $\RFC$, the systematic variations are performed based on the ratio
of the forward and central $\eta$ intervals
after each variation, to account for correlations between the two $\eta$ regions.
Thus, several of the effects discussed above, particularly the influence of the variations in the identification and isolation selection, partially cancel.

In the central $\eta$ region, the uncertainties at lower $\pT$ range from 18\% to 26\%,
and those at higher $\pT$ range from 8\% to 16\%.
In the forward $\eta$ region, the uncertainties at lower $\pT$ range from 20\% to 26\%,
and those at higher $\pT$ range from 13\% to 19\%.
For the yields, uncertainties for specific centrality, $\eta$ and $\pT$ ranges are provided in Table~\ref{yield_syst}.
For the ratio $\RFC$, the uncertainties at lower $\pT$ range from 8\% to 17\% and at higher $\pT$ from 6\% to 12\%.
Uncertainties for specific centrality, $\eta$ and $\pT$ ranges are provided in Table~\ref{yield_syst}.
For the ratios, uncertaities for specific centrality and $\pT$ ranges are provided in Table~\ref{ratio_syst}.

\section{Theoretical predictions}
\label{sect:theory}
$\jetphox$ 1.3 is used for
NLO pQCD calculations to compare with the
fully corrected measurements.  \jetphox\ was found to agree well (within 10--15\%) with
$\pbarp$ from the Tevatron~\cite{Abazov:2005wc,Aaltonen:2009ty}
and $\pp$ data from the LHC~\cite{Aad:2010sp,Aad:2011tw,Aad:2013zba}.
It provides access to a wide
range of existing PDF sets and performs
calculations for direct photon production as well as for photons
from fragmentation processes, both using an implementation of the
experimental isolation selection built into the calculations.  The
primary $\pp$ calculations shown in this work use the CTEQ6.6~\cite{Nadolsky:2008zw} proton PDF,
with no nuclear modification, and the BFG II fragmentation functions~\cite{Bourhis:1997yu}.
They require less than 6~\GeV\ isolation
energy in a cone of $\Riso = 0.3$ relative to the photon direction.
The effect of hadronization on the final cross sections was estimated using
the \PYTHIA\ 6.423 simulations to be 1\% or less, and is neglected in the
results shown here.
Scale uncertainties are estimated by varying the renormalization ($\mu_\mathrm{R}$),
factorization ($\mu_\mathrm{F}$) and fragmentation ($\mu_\mathrm{f}$) scales by a factor of two, relative to
the baseline result, $\mu_\mathrm{R} = \mu_\mathrm{F} = \mu_\mathrm{f} =
\pT^{\mathrm{photon}}$.  Two types of variations are performed, a
correlated variation of all three scales by a factor of two up and down,
as well as an independent variation of each scale up and down by a
factor of two, leaving the other two scales constant.  The envelope
covered by these variations is typically 12--18\%, varying with $\eta$ and
$\pT$.  PDF uncertainties are determined by varying the PDF fit parameters
according to 22 eigenvectors in the parameter space, and separately
keeping track of the upward and downward variations of the final
cross sections.  These uncertainties are generally less than 3\% for $\pT<100$~\GeV\
but increase to 6\% for $\pp$ for $\pT>140$~\GeV.
The impact of the uncertainty in the strong coupling constant
$\alpha_\mathrm{s} (M_{Z})$, $\Delta \alpha_\mathrm{s} = \pm 0.0012$,
was determined and found to be small.
For the yields it varies from $\pm (1 \ndash 2)\%$, decreasing with $\pT$.
For the ratio, it increases with $\pT$ from 0 to $2.5\%$.
These errors are not incorporated in the error bands shown.
The calculations were also performed with the MSTW2008 PDF~\cite{Martin:2009iq},
which yield cross sections about 6\% higher for $|\eta|<1.37$ for all calculated $\pT$ values.

To study nuclear effects, two additional calculations are performed.
The first reweights the contributions from up and down valence quarks to
account for the neutrons in the colliding lead nuclei, but with no
attempt at modeling the impact parameter dependence of the neutron
spatial distributions, e.g. due to a neutron skin.  This is a reasonable
first-order approximation for \PbPb\ (both with A=208) collisions using the standard PDF.
The other incorporates nuclear modifications to the nucleon parton distributions
using the EPS09~\cite{Eskola:2009uj} PDF set, which are $x$- and $Q^2$-dependent
modifications of the CTEQ 6.1 PDF, defined as ratios of the standard PDF
as a function of $x$ at a hardness scale $Q^2_0 = 1.69$~\GeV $^2$ and
evolved to the relevant $Q^2$ using standard DGLAP evolution.
The EPS09 modifications have their own set of 15 uncertainty eigenvectors, which are
used to evaluate 30 variations of the cross sections relative to the
default set, which are typically approximately 5\%, with only a small variation
in $\pT$.

\section{Results}
\begin{figure*}[t!]
\begin{center}
\includegraphics[width=0.95\textwidth]{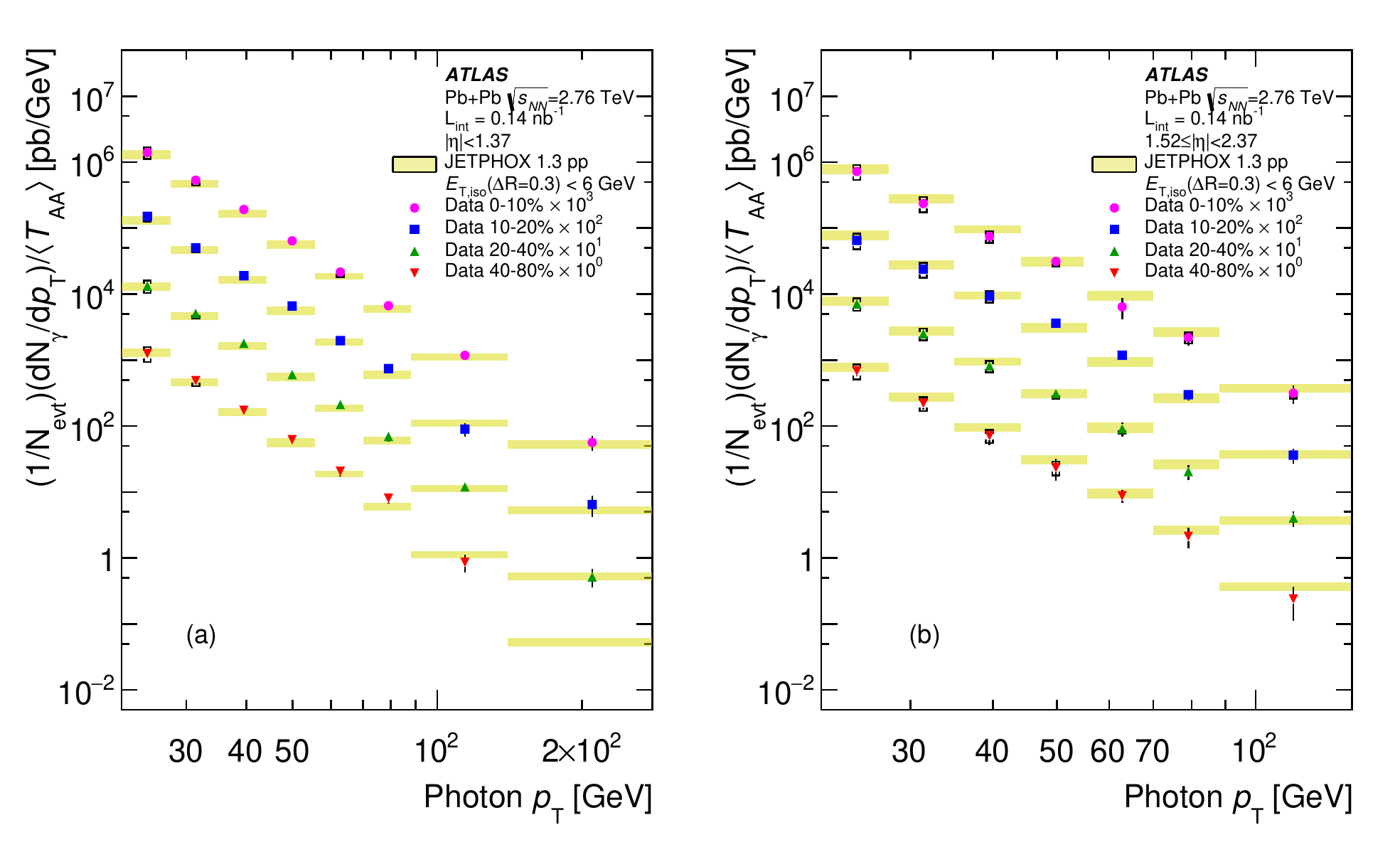}
\caption{  (Color online) Fully corrected yields of prompt photons in four centrality
  intervals as a function of $\pT$ in $|\eta|<1.37$ (a) and $1.52
  \leq |\eta|<2.37$ (b) using tight selection, isolation cone size
  $\Riso = 0.3$ and isolation transverse energy of less than 6~\GeV.
  \jetphox\ calculations, for proton-proton collisions and using the
  same isolation criterion, are shown by the yellow bands.
  Statistical uncertainties are shown by the error bars.  Systematic
  uncertainties on the photon yields shown by braces, which are
  smaller than the markers for some points.  The scale uncertainties
  due to $\meanTAA$ are tabulated for each bin in
  Table~\ref{table:centrality}.
\label{figure:yields}
}
\end{center}
\end{figure*}

\begin{sidewaystable}
\caption{
\label{yield_jetphox_table_eta0}
$\meanTAA$-scaled prompt photon yields compared with \jetphox\ 1.3
$pp$, for $|\eta|<1.37$ in four centrality intervals and for
\jetphox\ as a function of photon $\pT$.  For each value, the first
uncertainty is statistical and the second is systematic.  For
\jetphox\, the combined error is shown.  }
\footnotesize
\begin{center}
\sisetup{retain-zero-exponent = true}
\begin{tabular}{|r@{--}l|c|l|l|l|l|c|}
\hline
\multicolumn{3}{|c|}{}& \multicolumn{4}{c}{$dN/d\pT$ / $\meanTAA$ [pb/GeV] } & $d\sigma/d\pT$ [pb/GeV]\\
\hline
\multicolumn{2}{|c|}{$\pT$ [GeV]} & Scale & 40--80\% & 20--40\% & 10--20\% & 0--10\% & $\jetphox$ \\
\hline
\hline
22 & 28 & $10^{3}$ & $ 1.26 \pm 0.12 \pm 0.32 $ &  $ 1.32 \pm 0.06 \pm 0.29 $ &  $ 1.51 \pm 0.06 \pm 0.27 $ &  $ 1.40 \pm 0.06 \pm 0.29 $ & $1.31 ^{+0.20}_{-0.20} $ \\ [1ex]
28 & 35 & $10^{2}$ & $ 4.88 \pm 0.42 \pm 0.87 $ &  $ 5.09 \pm 0.25 \pm 0.82 $ &  $ 5.03 \pm 0.26 \pm 0.77 $ &  $ 5.33 \pm 0.25 \pm 0.91 $ & $4.70 ^{+0.65}_{-0.65} $ \\ [1ex]
35 & 44.1 & $10^{2}$ & $ 1.73 \pm 0.17 \pm 0.26 $ &  $ 1.79 \pm 0.09 \pm 0.23 $ &  $ 1.89 \pm 0.10 \pm 0.25 $ &  $ 1.92 \pm 0.09 \pm 0.27 $ & $1.66 ^{+0.22}_{-0.22} $ \\ [1ex]
44.1 & 55.6 & $10^{1}$ & $ 6.21 \pm 0.64 \pm 0.72 $ &  $ 6.01 \pm 0.40 \pm 0.69 $ &  $ 6.60 \pm 0.44 \pm 0.83 $ &  $ 6.42 \pm 0.40 \pm 0.96 $ & $5.66 ^{+0.85}_{-0.85} $ \\ [1ex]
55.6 & 70 & $10^{1}$ & $ 2.07 \pm 0.33 \pm 0.25 $ &  $ 2.12 \pm 0.19 \pm 0.24 $ &  $ 1.97 \pm 0.19 \pm 0.23 $ &  $ 2.16 \pm 0.21 \pm 0.34 $ & $1.88 ^{+0.22}_{-0.22} $ \\ [1ex]
70 & 88.2 & $10^{0}$ & $ 8.06 \pm 1.39 \pm 0.83 $ &  $ 6.96 \pm 1.11 \pm 0.83 $ &  $ 7.43 \pm 0.81 \pm 0.90 $ &  $ 6.66 \pm 0.81 \pm 0.98 $ & $6.05 ^{+0.84}_{-0.84} $ \\ [1ex]
88.2 & 140 & $10^{-1}$ & $ 8.60 \pm 2.59 \pm 0.87 $ &  $ 11.96 \pm 1.45 \pm 0.99 $ &  $ 8.99 \pm 2.09 \pm 1.08 $ &  $ 11.79 \pm 1.49 \pm 1.40 $ & $11.26 ^{+1.41}_{-1.39} $ \\ [1ex]
140 & 280 & $10^{-2}$ & &  $ 5.16 \pm 1.62 \pm 0.41 $ &  $ 6.47 \pm 2.29 \pm 0.65 $ &  $ 5.63 \pm 1.42 \pm 0.58 $ & $5.32 ^{+0.77}_{-0.74} $ \\ [1ex]
\hline
\hline
\end{tabular}
\end{center}
\end{sidewaystable}

\begin{sidewaystable}
\caption{
\label{yield_jetphox_table_eta1}
$\meanTAA$-scaled prompt photon yields compared with \jetphox\ 1.3 $pp$ for $1.52 \leq |\eta|<2.37$ in four centrality intervals and for \jetphox\ as a function of photon $\pT$.  For each value, the first uncertainty is statistical and the second is systematic.  For \jetphox\, the combined error is shown.
}
\footnotesize
\begin{center}
\sisetup{retain-zero-exponent = true}
\begin{tabular}{|r@{--}l|c|l|l|l|l|c|}
\hline
\multicolumn{3}{|c|}{}&\multicolumn{4}{c}{$dN/d\pT$ / $\meanTAA$ [pb/GeV] } & $d\sigma/d\pT$ [pb/GeV] \\
\hline
\multicolumn{2}{|c|}{$\pT$ [GeV]} & Scale & 40--80\% & 20--40\% & 10--20\% & 0--10\% & $\jetphox$ \\
\hline
\hline
22 & 28 & $10^{2}$ & $ 6.82 \pm 1.11 \pm 1.70 $ &  $ 7.08 \pm 0.56 \pm 1.49 $ &  $ 6.52 \pm 0.66 \pm 1.74 $ &  $ 7.22 \pm 0.55 \pm 1.82 $ & $7.90 ^{+1.33}_{-1.34} $ \\ [1ex]
28 & 35 & $10^{2}$ & $ 2.22 \pm 0.44 \pm 0.53 $ &  $ 2.50 \pm 0.24 \pm 0.50 $ &  $ 2.38 \pm 0.28 \pm 0.61 $ &  $ 2.36 \pm 0.24 \pm 0.61 $ & $2.80 ^{+0.45}_{-0.45} $ \\ [1ex]
35 & 44.1 & $10^{1}$ & $ 7.13 \pm 1.95 \pm 1.60 $ &  $ 8.13 \pm 1.14 \pm 1.64 $ &  $ 9.32 \pm 0.98 \pm 1.87 $ &  $ 7.48 \pm 1.39 \pm 1.53 $ & $9.62 ^{+1.35}_{-1.35} $ \\ [1ex]
44.1 & 55.6 & $10^{1}$ & $ 2.34 \pm 0.85 \pm 0.54 $ &  $ 3.10 \pm 0.41 \pm 0.50 $ &  $ 3.62 \pm 0.26 \pm 0.50 $ &  $ 3.13 \pm 0.28 \pm 0.49 $ & $3.13 ^{+0.52}_{-0.52} $ \\ [1ex]
55.6 & 70 & $10^{0}$ & $ 8.78 \pm 1.87 \pm 1.20 $ &  $ 9.08 \pm 2.16 \pm 1.40 $ &  $ 11.86 \pm 1.24 \pm 1.63 $ &  $ 6.41 \pm 2.25 \pm 0.88 $ & $9.56 ^{+1.69}_{-1.69} $ \\ [1ex]
70 & 88.2 & $10^{0}$ & $ 2.13 \pm 0.72 \pm 0.32 $ &  $ 2.04 \pm 0.54 \pm 0.27 $ &  $ 2.98 \pm 0.52 \pm 0.37 $ &  $ 2.19 \pm 0.54 \pm 0.42 $ & $2.68 ^{+0.45}_{-0.45} $ \\ [1ex]
88.2 & 140 & $10^{-1}$ & $ 2.39 \pm 1.26 \pm 0.35 $ &  $ 4.04 \pm 1.10 \pm 0.54 $ &  $ 3.61 \pm 0.95 \pm 0.46 $ &  $ 3.15 \pm 1.01 \pm 0.55 $ & $3.74 ^{+0.55}_{-0.55} $ \\ [1ex]
\hline
\hline
\end{tabular}
\end{center}
\end{sidewaystable}

The per-event differential photon yields are calculated according to Eq.~(\ref{eq:yield}).
These are then divided by $\meanTAA$ for comparison with the \jetphox\ calculations.
The results are shown as a function of $\pT$ in Fig.~\ref{figure:yields} and are tabulated
in Table~\ref{yield_jetphox_table_eta0} for $|\eta|<1.37$ and
in Table~\ref{yield_jetphox_table_eta1} for $1.52 \leq |\eta|<2.37$.
Each panel shows a single pseudorapidity interval, with four centrality intervals,
each scaled by a factor of 10 relative to each other.

\begin{figure*}[t!]
\begin{center}
\includegraphics[width=0.99\textwidth]{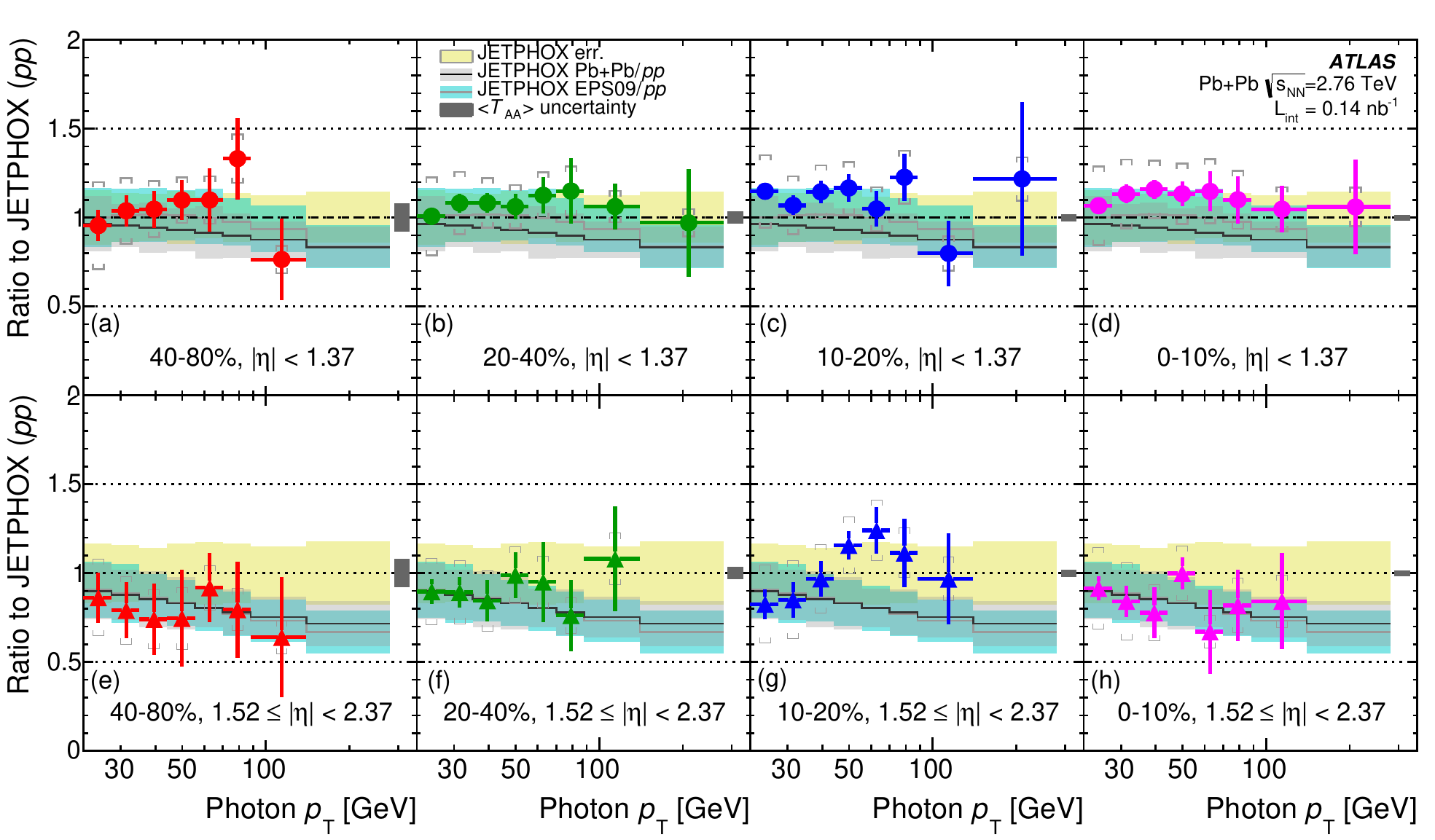}
\caption{ (Color online) Fully corrected normalized yields of prompt photons as a
  function of $\pT$ in $|\eta|<1.37$ ((a)-(d)) and $1.52\leq |\eta|<2.37$ ((e)-(h)) using
  tight photon selection, isolation cone size $\Riso = 0.3$ and
  isolation transverse energy of less than 6~\GeV, divided by
  \jetphox\ predictions for \pp\ collisions, which implement the same
  isolation selection.  The combined scale and PDF uncertainty on the
  \jetphox\ calculation is shown by the grey line with yellow area.  In addition two
  other \jetphox\ calculations are shown, also divided by the
  \pp\ results: \PbPb\ collisions with no nuclear modification
  (black line with grey area), and \PbPb\ collisions with EPS09 nuclear modifications
  (grey line with blue area).  Statistical uncertainties are shown by the bars.
  Systematic uncertainties on the photon yields are combined and shown
  by the upper and lower braces.  The scale uncertainties due only to
  $\meanTAA$ are tabulated for each bin in
  Table~\ref{table:centrality}.
\label{figure:scaledyields}
}
\end{center}
\end{figure*}

The ratios of the data to the \jetphox\ \pp\ predictions are shown in Fig.~\ref{figure:scaledyields}, and
are tabulated
in Table~\ref{ratio_jetphox_table_eta0} for $|\eta|<1.37$ and
in Table~\ref{ratio_jetphox_table_eta1} for $1.52 \leq |\eta|<2.37$.
In addition, the two other \jetphox\ calculations described in the previous section are
shown, also divided by the \pp\ results:
\PbPb\ collisions with no nuclear modification (black line),
and \PbPb\ collisions with EPS09 nuclear modifications (hatched blue area).
The combined scale and PDF uncertainty on the \jetphox\ calculations, calculated separately for each
configuration, are shown as shaded regions.
The data are found to agree well with the \jetphox\ \pp\ prediction
in all centrality and $\eta$ regions,
within the stated statistical and systematic uncertainties.
They are also consistent within uncertainties of the other physics scenarios as well.
Thus, the current data are not of sufficient precision to address nuclear PDF effects
quantitatively.
However, it should be noted that where the data are more precise,
in the central $\eta$ region, the PDF modifications implemented in EPS09
compensate for the suppression at higher $\pT$ seen in the $\PbPb$ calculations,
giving cross sections similar to the $\pp$ case.

\begin{table*}[b]
\caption{
\label{ratio_jetphox_table_eta0}
$\meanTAA$-scaled prompt photon yields divided by the cross section from $pp$ \jetphox\ 1.3, for $|\eta|<1.37$ in four centrality intervals as a function of photon $\pT$.
}
\footnotesize
\begin{center}
\begin{tabular}{|r@{--}l|c|c|c|c|c|}
\hline
\multicolumn{2}{|c|}{} & \multicolumn{5}{c|}{ $dN/d\pT / \meanTAA / d\sigma/d\pT$($\jetphox$)  }  \\
\hline
\multicolumn{2}{|c|}{$\pT$ [GeV]} & 40--80\% & 20--40\% & 10--20\% & 0--10\% & \jetphox\ \\
\hline
22 & 28 &  $0.95 \pm 0.09 \pm 0.24$ &  $1.01 \pm 0.05 \pm 0.22$ &  $1.15 \pm 0.04 \pm 0.20$ &  $1.07 \pm 0.04 \pm 0.22$ & $1^{+0.15}_{-0.15}$ \\ [1ex]
28 & 35 &  $1.04 \pm 0.09 \pm 0.18$ &  $1.08 \pm 0.05 \pm 0.17$ &  $1.07 \pm 0.06 \pm 0.16$ &  $1.13 \pm 0.05 \pm 0.19$ & $1^{+0.14}_{-0.14}$ \\ [1ex]
35 & 44.1 &  $1.05 \pm 0.11 \pm 0.16$ &  $1.08 \pm 0.06 \pm 0.14$ &  $1.14 \pm 0.06 \pm 0.15$ &  $1.16 \pm 0.05 \pm 0.16$ & $1^{+0.13}_{-0.13}$ \\ [1ex]
44.1 & 55.6 &  $1.10 \pm 0.11 \pm 0.13$ &  $1.06 \pm 0.07 \pm 0.12$ &  $1.17 \pm 0.08 \pm 0.15$ &  $1.13 \pm 0.07 \pm 0.17$ & $1^{+0.15}_{-0.15}$ \\ [1ex]
55.6 & 70 &  $1.10 \pm 0.18 \pm 0.13$ &  $1.13 \pm 0.10 \pm 0.13$ &  $1.05 \pm 0.10 \pm 0.12$ &  $1.15 \pm 0.11 \pm 0.18$ & $1^{+0.12}_{-0.12}$ \\ [1ex]
70 & 88.2 &  $1.33 \pm 0.23 \pm 0.14$ &  $1.15 \pm 0.18 \pm 0.14$ &  $1.23 \pm 0.13 \pm 0.15$ &  $1.10 \pm 0.13 \pm 0.16$ & $1^{+0.14}_{-0.14}$ \\ [1ex]
88.2 & 140 &  $0.76 \pm 0.23 \pm 0.08$ &  $1.06 \pm 0.13 \pm 0.09$ &  $0.80 \pm 0.19 \pm 0.10$ &  $1.05 \pm 0.13 \pm 0.12$ & $1^{+0.12}_{-0.12}$ \\ [1ex]
140 & 280 &  &  $0.97 \pm 0.30 \pm 0.08$ &  $1.22 \pm 0.43 \pm 0.12$ &  $1.06 \pm 0.27 \pm 0.11$ & $1^{+0.15}_{-0.14}$ \\ [1ex]
\hline
\hline
\end{tabular}
\end{center}
\end{table*}

\begin{table*}[b]
\caption{
\label{ratio_jetphox_table_eta1}
$\meanTAA$-scaled prompt photon yields divided by the cross section
from $pp$ \jetphox\ 1.3, for $1.52 \leq |\eta|<2.37$ in four
centrality intervals as a function of photon $\pT$.  } \footnotesize
\begin{center}
\begin{tabular}{|r@{--}l|c|c|c|c|c|}
\hline
\multicolumn{2}{|c|}{}& \multicolumn{5}{c|}{ $dN/d\pT / \meanTAA / d\sigma/d\pT$ ($\jetphox$) }   \\
\hline
\multicolumn{2}{|c|}{$\pT$ [GeV]} & 40--80\% & 20--40\% & 10--20\% & 0--10\% & $\jetphox$ \\
\hline
22 & 28 &  $0.86 \pm 0.14 \pm 0.22$ &  $0.90 \pm 0.07 \pm 0.19$ &  $0.83 \pm 0.08 \pm 0.22$ &  $0.91 \pm 0.07 \pm 0.23$ & $1^{+0.17}_{-0.17}$ \\ [1ex]
28 & 35 &  $0.79 \pm 0.16 \pm 0.19$ &  $0.89 \pm 0.09 \pm 0.18$ &  $0.85 \pm 0.10 \pm 0.22$ &  $0.84 \pm 0.09 \pm 0.22$ & $1^{+0.16}_{-0.16}$ \\ [1ex]
35 & 44.1 &  $0.74 \pm 0.20 \pm 0.17$ &  $0.84 \pm 0.12 \pm 0.17$ &  $0.97 \pm 0.10 \pm 0.19$ &  $0.78 \pm 0.14 \pm 0.16$ & $1^{+0.14}_{-0.14}$ \\ [1ex]
44.1 & 55.6 &  $0.75 \pm 0.27 \pm 0.17$ &  $0.99 \pm 0.13 \pm 0.16$ &  $1.16 \pm 0.08 \pm 0.16$ &  $1.00 \pm 0.09 \pm 0.15$ & $1^{+0.17}_{-0.17}$ \\ [1ex]
55.6 & 70 &  $0.92 \pm 0.20 \pm 0.13$ &  $0.95 \pm 0.23 \pm 0.15$ &  $1.24 \pm 0.13 \pm 0.17$ &  $0.67 \pm 0.24 \pm 0.09$ & $1^{+0.18}_{-0.18}$ \\ [1ex]
70 & 88.2 &  $0.80 \pm 0.27 \pm 0.12$ &  $0.76 \pm 0.20 \pm 0.10$ &  $1.11 \pm 0.19 \pm 0.14$ &  $0.82 \pm 0.20 \pm 0.16$ & $1^{+0.17}_{-0.17}$ \\ [1ex]
88.2 & 140 &  $0.64 \pm 0.34 \pm 0.09$ &  $1.08 \pm 0.29 \pm 0.15$ &  $0.97 \pm 0.26 \pm 0.12$ &  $0.84 \pm 0.27 \pm 0.15$ & $1^{+0.15}_{-0.15}$ \\ [1ex]
\hline
\hline
\end{tabular}
\end{center}
\end{table*}

\begin{figure*}[t!]
\begin{center}
\includegraphics[width=0.99\textwidth]{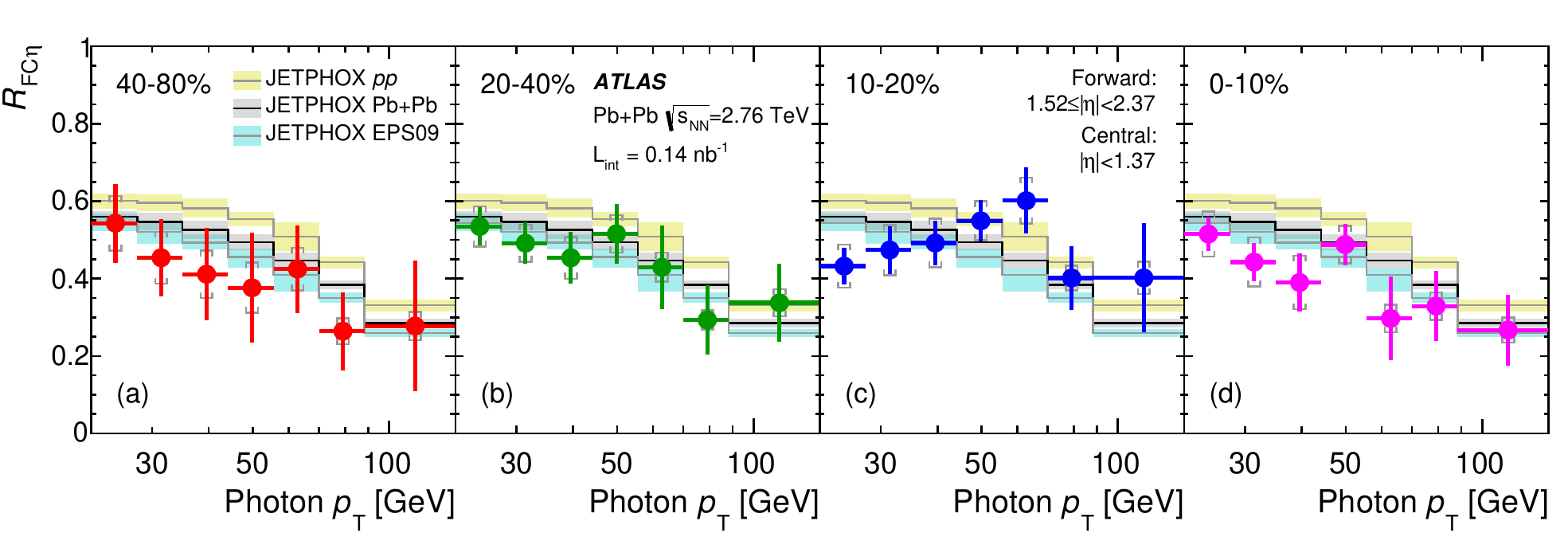}
\caption{  (Color online) Fully corrected yields of prompt photons as a function of
  $\pT$ in $1.52 \leq |\eta|<2.37$ divided by that measured in
  $|\eta|<1.37$ using the tight photon selection, isolation cone size
  \Riso\ = 0.3 and isolation transverse energy of 6~\GeV, for four centrality intervals ((a)-(d)).  The yield
  ratio is compared to $\jetphox$ 1.3 predictions that implement the
  same isolation selection, for three different configurations: for
  \pp\ collisions (grey line with yellow area), \PbPb\ collisions with no nuclear
  modification (black line with grey area), and \PbPb\ collisions with EPS09
  nuclear modifications (grey line with blue area).  Statistical uncertainties are
  shown by the bars.  Systematic uncertainties on the photon yields
  are combined and shown by the braces.
\label{figure:etaratio}
}
\end{center}
\end{figure*}

The ratios $\RFC$ of cross sections
between the forward and central $\eta$ intervals, are calculated as a function of $\pT$
for each centrality interval, and are shown in Fig.~\ref{figure:etaratio},
as well as tabulated in Table~\ref{table:etaratio_jetphox_table}.
Evaluation of these ratios leads to the cancellation of
several systematic effects on the efficiencies and
bin-by-bin correction factors,
mitigate the effect of the theoretical uncertainties,
and fully remove the uncertainty on $\meanTAA$.
The results are compared to \jetphox\ calculations for $\pp$ (yellow region), $\PbPb$ (black line with grey area)
and EPS09 nPDF (grey line with blue area).
It is clear that there is some sensitivity to the nuclear PDF,
primarily through the expected depletion of photon yields in the forward direction expected when including
the neutron PDF to match the isospin composition of the lead nuclei.
While the data are consistent with all three curves within the
statistical and systematic uncertainties, a slight preference for the calculations incorporating isospin effects is observed.

\begin{table*}[b]
\footnotesize
\begin{center}
\caption{
\label{table:etaratio_jetphox_table}
Results for $\RFC$, the prompt photon
yield in the forward $\eta$ region divided by that in the central $\eta$ region,
as a function of photon $\pT$ for four centrality bins ((a)-(d)).  \jetphox\ 1.3 $\pp$ calculations
are also provided.
}
\begin{tabular}{|r@{--}l|c|c|c|c|c|}
\hline
\multicolumn{2}{|c|}{}& \multicolumn{4}{c}{ $\RFC = dN/d\pT (1.52 \leq |\eta|<2.37)/ dN/d\pT (|\eta|<1.37)$ } &   \\
\hline
\multicolumn{2}{|c|}{$\pT$ [$\GeV$]} & 40--80\% & 20--40\% & 10--20\% & 0--10\% & $\jetphox$ \\
\hline
22 & 28 &  $0.54 \pm 0.10 \pm 0.07$ &  $0.53 \pm 0.05 \pm 0.05$ &  $0.43 \pm 0.05 \pm 0.06$ &  $0.52 \pm 0.04 \pm 0.06$ & $0.60 ^{+0.02}_{-0.02}$ \\  [1ex]
28 & 35 &  $0.45 \pm 0.10 \pm 0.07$ &  $0.49 \pm 0.05 \pm 0.05$ &  $0.47 \pm 0.06 \pm 0.06$ &  $0.44 \pm 0.05 \pm 0.06$ & $0.60 ^{+0.02}_{-0.02}$ \\ [1ex]
35 & 44.1 &  $0.41 \pm 0.12 \pm 0.06$ &  $0.45 \pm 0.07 \pm 0.05$ &  $0.49 \pm 0.06 \pm 0.06$ &  $0.39 \pm 0.07 \pm 0.07$ & $0.58 ^{+0.03}_{-0.02}$ \\ [1ex]
44.1 & 55.6 &  $0.38 \pm 0.14 \pm 0.06$ &  $0.52 \pm 0.08 \pm 0.05$ &  $0.55 \pm 0.05 \pm 0.05$ &  $0.49 \pm 0.05 \pm 0.05$ & $0.55 ^{+0.02}_{-0.02}$ \\ [1ex]
55.6 & 70 &  $0.42 \pm 0.11 \pm 0.06$ &  $0.43 \pm 0.11 \pm 0.04$ &  $0.60 \pm 0.09 \pm 0.06$ &  $0.30 \pm 0.11 \pm 0.04$ & $0.51 ^{+0.04}_{-0.04}$ \\ [1ex]
70 & 88.2 &  $0.26 \pm 0.10 \pm 0.03$ &  $0.29 \pm 0.09 \pm 0.02$ &  $0.40 \pm 0.08 \pm 0.03$ &  $0.33 \pm 0.09 \pm 0.04$ & $0.44 ^{+0.02}_{-0.02}$ \\ [1ex]
88.2 & 140 &  $0.28 \pm 0.17 \pm 0.04$ &  $0.34 \pm 0.10 \pm 0.03$ &  $0.40 \pm 0.14 \pm 0.04$ &  $0.27 \pm 0.09 \pm 0.03$ & $0.33 ^{+0.01}_{-0.02}$ \\ [1ex]
\hline
\hline
\end{tabular}
\end{center}
\end{table*}

\FloatBarrier

\section{Conclusions}
\label{sec:conclusion}

In this paper, measured yields of isolated prompt photons in 0.14 nb$^{-1}$ lead-lead
collisions recorded by the ATLAS detector at the LHC
have been presented as a function of collision centrality (in four intervals
from 40--80\% to 0--10\%),
in two pseudorapidity regions (\mbox{$|\eta|<1.37$} and \mbox{$1.52\leq|\eta|<2.37$})
and for photon transverse momenta in the range \mbox{$22 \leq \pT < 280$~\GeV }.
Photons were reconstructed using the large-acceptance, longitudinally segmented electromagnetic calorimeter,
after an event-by-event subtraction of the average underlying event in each calorimeter layer in
small $\Delta\eta$ intervals.
Backgrounds stemming from neutral hadrons in jets are suppressed by a tight shower shape selection and
by requiring no more than 6~\GeV\ transverse energy in a cone of size $\DR=0.3$ around each photon.
The residual hadronic background is determined using a double sideband method, and the remaining signal is
corrected for efficiency
and resolution, as well as electron contamination,
to arrive at the per-event yield of photons as a function of $\pT$, in each $\eta$ and centrality
interval.
After scaling the yields by the mean nuclear thickness $\meanTAA$,
the $\pT$ spectrum in each $\eta$ and centrality interval is found to agree,
within statistical and systematic uncertainties, with next-to-leading-order perturbative
QCD calculations of proton-proton collisions.
The data are also compared with calculations that assume the isospin of \PbPb\ collisions, as well as
the calculations for \PbPb\ using the EPS09 nuclear modifications of the proton parton distribution functions.
The ratios of the forward yields to those near midrapidity ($\RFC$) are also shown, and are compared to the
corresponding ratios from $\jetphox$.
The present data are unable to distinguish between the three scenarios.
However, the overall consistency of the measured yields with $\jetphox$ expectations for all centrality intervals
demonstrates that photon yields in heavy ion collisions scale as expected with the mean nuclear thickness.
This provides further support for the interpretation of the clear modification of jet yields
in \PbPb\ collisions as a function of centrality, relative
to those measured in proton-proton collisions, as stemming from
energy loss in the hot, dense medium~\cite{Aad:2014bxa}.

\section*{Acknowledgments}



We thank CERN for the very successful operation of the LHC, as well as the
support staff from our institutions without whom ATLAS could not be
operated efficiently.

We acknowledge the support of ANPCyT, Argentina; YerPhI, Armenia; ARC, Australia; BMWFW and FWF, Austria; ANAS, Azerbaijan; SSTC, Belarus; CNPq and FAPESP, Brazil; NSERC, NRC and CFI, Canada; CERN; CONICYT, Chile; CAS, MOST and NSFC, China; COLCIENCIAS, Colombia; MSMT CR, MPO CR and VSC CR, Czech Republic; DNRF, DNSRC and Lundbeck Foundation, Denmark; IN2P3-CNRS, CEA-DSM/IRFU, France; GNSF, Georgia; BMBF, HGF, and MPG, Germany; GSRT, Greece; RGC, Hong Kong SAR, China; ISF, I-CORE and Benoziyo Center, Israel; INFN, Italy; MEXT and JSPS, Japan; CNRST, Morocco; FOM and NWO, Netherlands; RCN, Norway; MNiSW and NCN, Poland; FCT, Portugal; MNE/IFA, Romania; MES of Russia and NRC KI, Russian Federation; JINR; MESTD, Serbia; MSSR, Slovakia; ARRS and MIZ\v{S}, Slovenia; DST/NRF, South Africa; MINECO, Spain; SRC and Wallenberg Foundation, Sweden; SERI, SNSF and Cantons of Bern and Geneva, Switzerland; MOST, Taiwan; TAEK, Turkey; STFC, United Kingdom; DOE and NSF, United States of America. In addition, individual groups and members have received support from BCKDF, the Canada Council, CANARIE, CRC, Compute Canada, FQRNT, and the Ontario Innovation Trust, Canada; EPLANET, ERC, FP7, Horizon 2020 and Marie Skłodowska-Curie Actions, European Union; Investissements d'Avenir Labex and Idex, ANR, Region Auvergne and Fondation Partager le Savoir, France; DFG and AvH Foundation, Germany; Herakleitos, Thales and Aristeia programmes co-financed by EU-ESF and the Greek NSRF; BSF, GIF and Minerva, Israel; BRF, Norway; the Royal Society and Leverhulme Trust, United Kingdom.

The crucial computing support from all WLCG partners is acknowledged
gratefully, in particular from CERN and the ATLAS Tier-1 facilities at
TRIUMF (Canada), NDGF (Denmark, Norway, Sweden), CC-IN2P3 (France),
KIT/GridKA (Germany), INFN-CNAF (Italy), NL-T1 (Netherlands), PIC (Spain),
ASGC (Taiwan), RAL (UK) and BNL (USA) and in the Tier-2 facilities
worldwide.


\providecommand{\href}[2]{#2}\begingroup\raggedright\endgroup


\clearpage

\newpage
\begin{flushleft}
{\Large The ATLAS Collaboration}

\bigskip

G.~Aad$^{\rm 84}$,
B.~Abbott$^{\rm 112}$,
J.~Abdallah$^{\rm 152}$,
S.~Abdel~Khalek$^{\rm 116}$,
O.~Abdinov$^{\rm 11}$,
R.~Aben$^{\rm 106}$,
B.~Abi$^{\rm 113}$,
M.~Abolins$^{\rm 89}$,
O.S.~AbouZeid$^{\rm 159}$,
H.~Abramowicz$^{\rm 154}$
% ...
\bigskip
\\
% ...
$^{11}$ Institute of Physics, Azerbaijan Academy of Sciences, Baku, Azerbaijan\\
% ...
$^{84}$ CPPM, Aix-Marseille Universit{\'e} and CNRS/IN2P3, Marseille, France\\
% ...
$^{89}$ Department of Physics and Astronomy, Michigan State University, East Lansing MI, United States of America\\
% ...
$^{106}$ Nikhef National Institute for Subatomic Physics and University of Amsterdam, Amsterdam, Netherlands\\
% ...
$^{112}$ Homer L. Dodge Department of Physics and Astronomy, University of Oklahoma, Norman OK, United States of America\\
$^{113}$ Department of Physics, Oklahoma State University, Stillwater OK, United States of America\\
% ...
$^{116}$ LAL, Universit{\'e} Paris-Sud and CNRS/IN2P3, Orsay, France\\
% ...
$^{152}$ Institute of Physics, Academia Sinica, Taipei, Taiwan\\
% ...
$^{154}$ Raymond and Beverly Sackler School of Physics and Astronomy, Tel Aviv University, Tel Aviv, Israel\\
% ...
$^{159}$ Department of Physics, University of Toronto, Toronto ON, Canada\\
% ...
\end{flushleft}

\end{document}